# POSIX-based Operating System in the environment of NVM/SCM memory

Vyacheslav Dubeyko, Cyril Guyot, Luis Cargnini, Adam Manzanares

Western Digital Research

{Vyacheslav.Dubeyko, Cyril.Guyot, Luis.Cargnini, Adam.Manzanares}@wdc.com

*Abstract*—**Modern Operating Systems are typically POSIX-compliant. The system calls are the fundamental layer of interaction between user-space applications and the OS kernel and its implementation of fundamental abstractions and primitives used in modern computing. The next generation of NVM/SCM memory raises critical questions about the efficiency of modern OS architecture. This paper investigates how the POSIX API drives performance for a system with NVM/SCM memory. We show that OS and metadata related system calls represent the most important area of optimization. However, the synchronization related system calls (poll(), futex(), wait4()) are the most time-consuming overhead that even a RAMdisk platform fails to eliminate. Attempting to preserve the POSIX-based approach will likely result in fundamental inefficiencies for any future applications of NVM/SCM memory.**

## I. INTRODUCTION

The next generation of Non-Volatile Memories (NVM) also known as Storage Class Memories (SCM) raise critical questions about the efficiency of modern OS architecture (for example, Linux OS). What should the OS architecture be in the case of NVM/SCM memory? How can file system architecture unleash the potential of NVM/SCM memory? These are the important questions that need to be answered in the near future. But where shall we start? Our answer is to investigate the system calls. Modern Operating Systems are typically POSIX-compliant. POSIX defines the application programming interface (API), along with command line shells and utility interfaces. The libc and libstdc++ are the final boundaries for any user-space applications. These libraries implement a set of primitives that interact with kernel-space via system calls in order to delegate some low-level tasks to the kernel. In this work, we attempt to understand the interaction dynamics between user space and kernel space for several typical user-space applications (software installation, diff viewer, Firefox, OpenOffice Calc, compilation). We studied the strace output for several platforms (AMD E-450, i7-3630QM, XEON E5-

2620 v2) with the Ubuntu Linux distribution (3.13.0-24, 3.13.0-95, 4.2.0-42 kernel versions) on different storage devices HDDs (5200 RPM SATA 3.0, 15015 RPM SAS), SSD (SATA 3.1, PCIe) and RAMdisk while using different file systems (ext2, ext4, XFS, tmpfs). As a result, we categorized applications into three main types: OS intensive, metadata intensive and mixed (OS + metadata). Every application calls some primary set of system calls with frequencies that dominate against the rest of system calls that are called by application. This primary set of system calls derives from the nature of the application and decides its total execution time. Metadata-related system calls appear to be a more important optimization factor for the case of NVM/SCM memory than the user data related system calls. Most critical above all is the synchronization overhead. This overhead remains practically the same even for the RAMdisk based platform. Also synchronization overhead is the most unpredictable and variable component of the total execution time.

## II. BACKGROUND & MOTIVATION

The inventors of the first computer systems had to make a series of compromises in the face of the available technologies at the time. One of the most critical problem was persistent storage technology. If the researchers had had a fast, persistent, large, byte-addressable memory the whole computing paradigm would most likely have been dramatically different. Nevertheless, we still lack the wide availability of such an "ideal" memory. However, technological progress is potentially turning this "dream" into reality (PCM, MRAM, ReRAM, NRAM and so on). Those technologies are however still imperfect: high BER (Bit Error Rate), low endurance and mediocre latency. Even current state of the art of NVM/SCM memory reveals the deep deficiency of modern OS technologies. The nature of NVM/SCM memory reveals the contradiction with POSIX-based OS. On the one hand, the NVM/SCM memory is byte-addressable and reasonably fast, albeit not quite as fast as DRAM yet. On the other hand, NVM/SCM memory has: (1) a persistent nature, (2) high BER (Bit Error Rate), (3) lackluster endurance. If some controller hides NVM/SCM memory, like a storage device does, then the controller's overhead essentially neutralizes the advantages of NVM/SCM memory. Conversely, direct byte-addressable accesses to NVM/SCM memory cannot be used in current POSIX-based OSes. Nowadays, the main research efforts (Aerie [1], BAFS [2], BPFS [3], NOVA [4], PMFS [5], SCMFS [6], DAX [8]) are focused on attempts to modify the file system and/or block layer of POSIX-based OS when operating in an environment with NVM/SCM memory. The key motivation of the mentioned research works [1 - 8] was the understanding that file system and block layer are responsible for interaction with a persistent storage device. However, how does the POSIX API correspond to the case of NVM/SCM memory? How deep is our understanding of the generic nature of modern POSIX-based OS? This research provides an experimental basis to answer those questions. The system calls were selected as the basis for this research. The system calls' frequency of calls and consumed time in different type of applications are able to provide a vision of efficiency of POSIX-based OSes in the NVM/SCM memory environment. The motivation of this research is to realize the bottlenecks of POSIX model and to elaborate the vision of critical points that can be a basis for a new model of memory-centric non-POSIX OSes.

## III. RELATED WORKS

Volos et al. [1] suggested the Aerie's architecture that exposes file-system data stored in SCM directly to user-mode programs. Applications link to a file-system library that provides local access to data and communicates with a service for coordination. The OS kernel provides only coarse-grained allocation and protection, and most functionality is distributed to client programs. Son et al. [2] proposed I/O stack consists of the byte-capable interface and a user-level file system (BAFS) supporting byte-addressable I/O. The byte-capable interface eliminates VFS layer, page cache, block and SCSI layer in the traditional Linux I/O stack. It allows the file system to perform byte-addressable I/O and does not follow the design of existing block device driver based on block I/O. Suggested interface provides POSIX interface the file system and removes block-level optimizations such as I/O scheduler, asynchronous I/O, and block I/O. Condit et al. [3] suggested a file system (BPFS) and a hardware architecture that are designed around the properties of persistent, byte-addressable memory. The BPFS uses a technique called short-circuit shadow paging to provide atomic, fine-grained updates to persistent storage. Short-circuit shadow paging allows BPFS to use copy-on-write at fine granularity, atomically committing small changes at any level of the file system tree. BPFS can often avoid copies altogether, writing updates in place without sacrificing reliability. Xu et al. [4] presented NOVA, a file system designed to work on hybrid memory systems with trying to provide consistency guarantees. NOVA keeps log and file data in NVMM and builds radix trees in DRAM to quickly perform search operations. The leaves of the radix tree point to entries in the log which in turn point to file data. Each inode in NOVA has its own log, allowing concurrent updates across files without synchronization. To atomically write data to a log, NOVA first appends data to the log and then atomically updates the log tail to commit the updates, thus avoiding both the duplicate writes overhead of journaling file systems and the cascading update costs of shadow paging systems. NOVA uses a linked list of 4 KB NVM pages to hold the log and stores the next page pointer in the end of each log page. The inode logs in NOVA do not contain file data. Instead, NOVA uses copy-on-write for modified pages and appends metadata about the write to the log. The metadata describe the update and point to the data pages. Dulloor et al. [5] implemented PMFS, a light-weight POSIX file system that exploits PM's byte-addressability to avoid overheads of block-oriented storage and enable direct PM access by applications (with memory-mapped I/O). PMFS exploits the processor's paging and memory ordering features for optimizations such as fine-grained logging (for consistency) and transparent large page support (for faster memory-mapped I/O). To provide strong consistency guarantees, PMFS requires only a simple hardware primitive that provides software enforceable guarantees of durability and ordering of stores to PM. Finally, PMFS uses the processor's existing features to protect PM from stray writes, thereby improving reliability. Wu et al. [6] propose SCMFS file system which is implemented on the virtual address space. SCMFS utilizes the existing memory management module in the operating system to do the block management and keep the space always contiguous for each file. Wu et al. assume that the storage device, SCM, is directly attached to CPU, and there is a way for firmware/software to distinguish SCM from the other volatile memories. This assumption allows the file systems be able to access the data on SCM in the same way as normal RAM. With this assumption, it was utilized the existing memory management module in the operating system to manage the space on the storage class memory.

## IV. METHODOLOGY

### A. Experimental Setup

The key goal of the research was to detect in well-defined experimental setups the most critical factors that can significantly affect the efficiency of NVM/SCM memory in POSIX-based OS [9] environment.

| | | | Platform#1 | Platform#2 | Platform#3 | Platform#4 | Platform#5 | Platform#6 |
|---|---|---|---|---|---|---|---|---|
| Hardware | CPU | Type | AMD E-450 1650 MHz | AMD E-450 1650 MHz | i7-3630QM 2.40 GHz | Xeon E5-2620 v2 2.10 GHz | Xeon E5-2620 v2 2.10 GHz | Xeon E5-2620 v2 2.10 GHz |
| | | Cores | 2 cores | 2 cores | 8 cores | 24 cores | 24 cores | 24 cores |
| | | Cache | L1 32 KB, L2 512 KB | L1 32 KB, L2 512 KB | L1 32 KB, L2 256 KB, L3 6 MB | L1 32 KB, L2 256 KB, L3 15 MB | L1 32 KB, L2 256 KB, L3 15 MB | L1 32 KB, L2 256 KB, L3 15 MB |
| | Memory | Type | SODIMM DDR3 | SODIMM DDR3 | SODIMM DDR3 1600 MHz | DIMM DDR3 1866 MHz | DIMM DDR3 1866 MHz | DIMM DDR3 1866 MHz |
| | | Size | 8GiB | 8GiB | 24 GiB | 16GiB | 16GiB | 16GiB |
| | Persistent storage | Type | HDD | SSD | SSD | SSD | SSD | DRAM |
| | | RPM | 5200 RPM | - | - | 15015 RPM | - | - |
| | | Size | 2 TB | 500 GB | 480 GB | 73.4 GB | 128 GB | 16 GB |
| | | Protocol | SATA 3.0 6 Gb/s | SATA 3.1 6 Gb/s | SATA 3.1 6 Gb/s | SAS | PCIe, SATA 3.1 6 Gb/s | - |
| Software | | OS | Linux: 3.13.0-24 | Linux: 3.13.0-24 | Linux: 3.13.0-95 | Linux: 3.13.0-24 | Linux: 3.13.0-24 | Linux: 4.2.0-42 |
| | | File system | EXT2 | EXT4 | XFS | EXT2 | XFS | TMPFS |

Table 1: Hardware platforms.

As a result, the first important point was to define a set of platforms that would allow us to distinguish qualitative thresholds of the modern computing paradigm. The most critical artifacts of the computing paradigm are: (1) CPU, (2) DRAM, (3) storage device, (4) file system. Table 1 contains description of all hardware platforms used in the experiments. We selected several CPUs with: (1) different architectures (AMD E-450, XEON E5-2620, i7-3630QM), (2) one (AMD, i7) and two (XEON) physical sockets; (3) various core numbers (2 − 24 cores); (4) various L1/L2/L3 cache sizes. We varied the DRAM size from 8 GB up to 24 GB. The most important constraint of this research was the selection of a representative set of persistent storage devices. We used several HDDs (5200 RPM SATA 3.0, 15000 RPM SAS) and SSDs (SATA 3.1, PCIe). The RAMdisk [7] was chosen as a special case that approximates an "ideal" environment in order to estimate the efficiency of POSIX-based OS in an environment with fast persistent memory. The ext2, ext4 and xfs file system were selected because of their support for Direct Access (DAX) [8]. The tmpfs file system was used for the case of RAMdisk based platform.

### B. Use Cases

A computer system's goal is to create an environment for accessing, transforming and storing a user's data. Modern computing paradigm uses an application/process as the environment within which data is accessed through the OS namespace. From the user's perspective, only the GUI/UI/window/console remains visible amongst all the diverse technologies implemented by the OS. Finally, the system performance is driven by the time that the end user needs to wait for a request to be processed. We needed to choose such use-cases that exhibit: (1) critical bottlenecks of the whole system; (2) potential directions for performance improvements. As a result, we analyzed the following use-cases (Table 2): (1) software installation, (2) diff viewer, (3) internet browser, (4) OpenOffice, (5) compilation. These use-cases were selected as typical cases of end users usage scenarios in a desktop system because every user: (1) installs applications, (2) searches files and traverses folders, (3) uses an internet browser, (4) processes data in a variety of data processing tools, (5) uses complex mixed workloads that can be simulated by the compilation case. Software installation was expected to be dominated by write operations, the diff viewer use case by metadata operations, the internet browser by operations on lots of small files, OpenOffice to be dominated by read operations, and compilation a mixed case of read/write/metadata operations. The test workflow included: (1) start OS, (2) synchronize data on disk with memory, (3) drop all caches, (4) run use-case with strace logging, and, (5) optional window termination. The goal of the research was: (1) evaluate application profile – what system calls are dominant in that specific use-case; (2) evaluate the most time-consuming system calls in the use-case; (3) detect changes in time consumption on different hardware platforms for different use-cases.

| Use-case | | Description |
|---|---|---|
| Software installation | Legend | (1) Start OS; (2) Drop all caches; (3) Start Ubuntu software center; (4) Install application; (5) Close Ubuntu software center. |
| | Command line | sync<br>echo 3 > /proc/sys/vm/drop_caches<br>strace -f -c /usr/bin/python /usr/bin/software-center 2> ./install.summary |
| Meld (diff viewer) | Legend | (1) Start OS; (2) Drop all caches; (3) Start Meld; (4) Select comparing folders; (5) Compare two folders and merge content of one file; (5) Close Meld. |
| | Command line | sync<br>echo 3 > /proc/sys/vm/drop_caches<br>strace -f -c meld 2> ./meld.summary |
| Firefox | Legend | (1) Start OS; (2) Drop all caches; (3) Start Firefox; (4) Try to find something into google.com, follow by any link; (5) Close Firefox. |
| | Command line | sync<br>echo 3 > /proc/sys/vm/drop_caches<br>strace -f -c /usr/lib/firefox/firefox 2> ./firefox.summary |
| OpenOffice Calc | Legend | (1) Start OS; (2) Drop all caches; (3) Start OpenOffice Calc; (4) Fill the worksheet by data, store file; (5) Close OpenOffice Calc. |
| | Command line | sync<br>echo 3 > /proc/sys/vm/drop_caches<br>strace -f -c /usr/bin/soffice --calc 2> ./excel.summary |
| Compilation | Legend | (1) Start OS; (2) Drop all caches; (3) Start autoreconf utility in F2FS tools source folder; (4) Drop all caches; (5) Start configure script in F2FS tools source folder; (6) Drop all caches; (7) Start make utility in F2FS tools source folder. |
| | Command line | sync<br>echo 3 > /proc/sys/vm/drop_caches<br>strace -f -c autoreconf --install 2> ./autoreconf.summary<br>sync<br>echo 3 > /proc/sys/vm/drop_caches<br>strace -f -c ./configure 2> ./configure.summary<br>sync<br>echo 3 > /proc/sys/vm/drop_caches<br>strace -f -c ./make 2> ./make.summary |

Table 2: Use cases.

### C. Research Tool

The activity of any modern single- or multi-threaded application is split between user-space and kernel-space. Generally speaking, the total execution time of an application includes user-space time and kernel-space time. The kernel-space time is the aggregated time that all application's threads spend in system calls. Any system call (or kernel call) is a request to the Operating System made by a process/thread for a service performed by the kernel via POSIX API. We can summarize this time in the following formula:

$$ExecutionTime = UserTime$$
$$+ \sum_{k=1}^{threads} \sum_{j=1}^{syscalls} \sum_{i=1}^{T} \left( frequency_i * time_{(i,syscall_{j,k})} \right)$$

Generally speaking, the system calls can be seen as the vehicle by which user-space threads access NVM/SCM memory. If the POSIX API is used with NVM/SCM memory then system calls can significantly affect the performance of a system with this new type of persistent memory. As a result, it is possible to assume that the system calls will be the critical factor determining the whole system performance in the environment of NVM/SCM memory. It is possible to achieve a qualitative understanding of the nature of a given use case by: (1) registering all the system calls that are called in an application, (2) determining the frequency for all registered system calls, (3) measuring the system calls' latency/timing, (4) determining the variability of system calls' timing in the application. Clearly understanding the application profile helps understand how efficient POSIX-based OSes will be when used in conjunction with NVM/SCM memory. To that end, we used as main measurement tool the strace utility. It intercepts and records the system calls which are called by a process and the signals which are received by that same process. Table 2 shows how to use the strace utility to record the profile of a given use-case.

## D. System Calls Classification

The POSIX API represents the architecture and fundamental concepts of modern OSes. Cornerstone to modern POSIX-based OS is the process/thread and file concepts. These fundamental concepts build the OS architecture. The process and file concepts reveal the principal dichotomy of an object existing in volatile space (process/thread) and in non-volatile, persistent space (file). As a result, the whole POSIX API is split between two principal subsets: (1) system calls that manage objects in volatile memory (OS-oriented system calls), (2) system calls that manage persistent objects. The file abstraction introduces the concept of "infinite" byte-stream that is limited only by the available free space of a file system volume. An implementation of file abstraction needs two principal types of information: (1) the user data that makes up the file, (2) the metadata that describes the location and attributes of the user data. The POSIX API provides special system calls for accessing and modifying both user data and metadata. We are able to separate the POSIX API along three fundamental classes: (1) OS-related system calls, (2) Metadata-related system calls, (3) User data-related system calls. Table 3 shows how system calls can be separated along those three classes. The suggested classification provides a convenient basis to analyze the efficiency of POSIX-based OSes in a NVM/SCM environment. Anybody can understand the modern computing paradigm like two poles abstraction: (1) very fast CPU core, (2) slow persistent storage.

| OS related syscalls | |
|---|---|
| Process/Thread related syscalls | execve(), clone(), prctl(), arch_prctl(), set_tid_address(), gettid(), getrlimit(), setrlimit(), prlimit(), getrusage(), getpriority(), setpriority(), getresuid(), getresgid(), getuid(), geteuid(), getgid(), getegid(), unshare(), sched_setaffinity(), sched_getaffinity(), _exit(), Exit(), exit_group(), uname(), sysinfo(), clock_getres(), clock_gettime(), clock_settime(), vfork(), getpid(), getppid(), getpgrp() |
| Memory related syscalls | brk(), mprotect(), madvise(), shmget(), shmat(), shmdt(), shmctl() |
| Synchronization related syscalls | get_robust_list(), set_robust_list(), futex(), wait3(), wait4(), nanosleep(), poll(), ppoll(), epoll_create(), epoll_create1(), epoll_ctl(), epoll_wait(), epoll_pwait(), eventfd(), eventfd2(), select(), pselect(), inotify_init(), inotify_init1(), inotify_add_watch(), inotify_rm_watch() |
| Signal related syscalls | rt_sigaction(), rt_sigprocmask(), rt_sigreturn(), sigaltstack(), kill(), tkill(), tgkill(), restart_syscall() |
| Socket related syscalls | socket(), socketpair(), connect(), pipe(), pipe2(), getpeername(), getsockname(), getsockopt(), setsockopt(), bind(), recv(), recvfrom(), recvmsg(), send(), sendto(), sendmsg(), sendmmsg(), shutdown() |
| File object related syscalls | fcntl(), posix_fadvise(), fadvise64(), umask(), dup(), dup2(), dup3() |
| Metadata related syscalls | |
| | access(), close(), stat(), fstat(), lstat(), fstatat(), statfs(), fstatfs(), getdents(), getdents64(), lseek(), open(), openat(), creat(), ioctl(), unlink(), unlinkat(), fallocate(), rename(), renameat(), renameat2(), chmod(), fchmod(), fchmodat(), mkdir(), mkdirat(), symlink(), symlinkat(), quotactl(), getcwd(), rmdir(), chdir(), chown(), lgetxattr() |
| User data related syscalls | |
| | mmap(), munmap(), read(), readahead(), readlink(), readlinkat(), readv(), pread(), preadv(), write(), writev(), pwrite, pwrite(), fsync(), fdatasync(), truncate(), ftruncate() |

Table 3: System calls classification.

Generally speaking, modern POSIX-based OSes joins these two far-distant poles into efficient unity by means of preemptive multitasking and file system's page cache. The goal of suggested classification is to figure out the possible behavior and efficiency of POSIX-based OS in the environment of fast, persistent, and byte-addressable memory.

## V. Use Cases Analysis

### A. Software Installation

The strace output revealed that the Ubuntu Software Center is a multi-threaded application with 69 threads. Table 4 shows the detected number and aggregated frequencies of different system call classes. The experimental results clearly show that metadata related system calls are the most frequently called type of system calls in the software installation use-case.

| System call Class | | System calls Total | Aggregated Frequency | Execution Time (%) |
|---|---|---|---|---|
| | Process/Thread system calls | 10 | 26084 | 0.24% |
| | Memory system calls | 6 | 2674 | 0.07% |
| | Synchronization system calls | 7 | 108810 | 52.94% |
| | Signal system calls | 4 | 1112 | 27.23% |
| | Socket system calls | 9 | 30886 | 0.29% |
| | File object system calls | 4 | 17894 | 0.18% |
| OS related | | 40 | 187460 | 80.95% |
| Metadata related | | 19 | 1690530 | 17.72% |
| User data related | | 9 | 38669 | 1.29% |

Table 4: Total number, aggregated frequency of system calls and percentage of total execution time for the software installation use-case.

The whole application budget (total number of calls) is distributed between: (1) OS related system calls – 9.67%, (2) metadata related system calls – 88.16%, (3) user data related system calls – 2.01%. Finally, we found that the most frequent system calls are: (1) stat() – get file status – 87%, (2) poll() – wait for some event on a file descriptor – 5% (see Figure 1). Table 4 shows the distribution of total execution time between different system call classes. One conclusion is that the dominant components of the application activity are threads synchronization (52.94%) and signals processing (27.23%). However, Figure 1 makes clearly visible that the most greedy consumers of execution time are poll(), futex() and restart_syscall() system calls. The restart_syscall() has very low frequency and very heavy time consumption. This implies that a thread calling the restart_syscall() spends most of its time in sleep state. Moreover, the restart_syscall() usually appears at the end of application's lifecycle. Therefore the whole time consumed by the restart_syscall() can be completely excluded as "sleep time". Building the histogram for the case of poll() system call, for example (see Figure 2), one discovers that the timing of single system call fluctuates from 0.000018 seconds up to 213 seconds. Generally speaking, one can see that the range of timing between 0.00006 seconds up to 213 seconds is solely the sleep time of the application's threads. Excluding the sleep time changes the whole picture and the whole application's execution time budget ends up between: (1) OS related system calls – 18%, (2) metadata related system calls – 77.18%, (3) user data related system calls – 4.8%. Actually, it is possible to see that software installation use-case is a purely metadata intensive application. And the most time-consuming system calls are: (1) stat() – get file status – 76.9%, (2) poll() – wait for some event on a file descriptor – 16.1%.

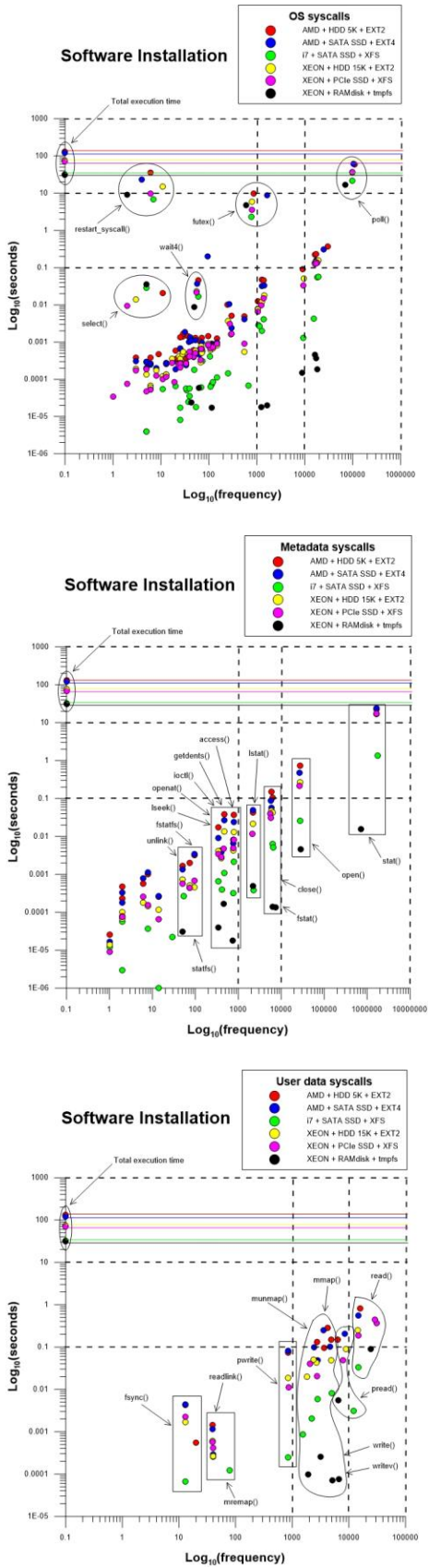

Figure 1: Software installation use-case.

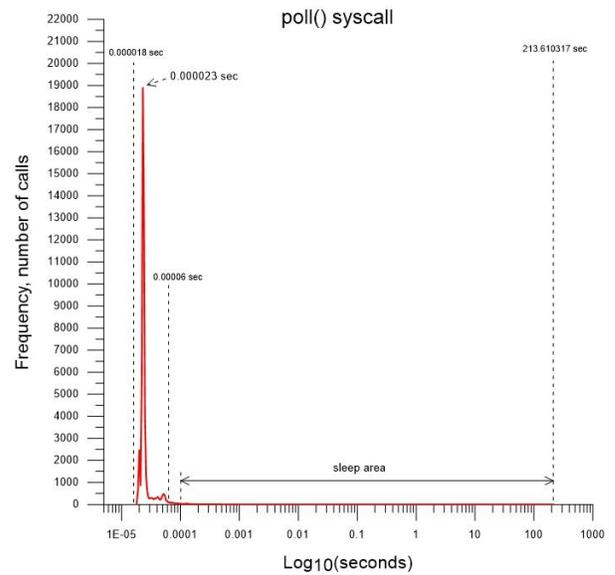

Figure 2: poll() system call histogram.

The comparison of results for different platforms (see Figure 1) shows very interesting and unexpected results. The read(), pread(), readlink() system calls are not significantly improved in the case of a RAMdisk based platform. Conversely, write() and writev() system calls are much more efficient. Metadata-related system calls are executed faster in the case of the RAMdisk-based platform. However, poll(), futex(), restart_syscall(), wait4(), select() system calls don't show any real improvements in that case. The software installation case is slightly faster on the RAMdisk-based platform but the total execution time is not significantly lower.

### B. Diff Viewer

The strace output revealed that Meld is a multi-threaded application with 55 threads.

| | System call Class | System calls Total | Aggregated Frequency | Execution Time (%) |
|---|---|---|---|---|
| | Process/Thread system calls | 9 | 79 | 0.014% |
| | Memory system calls | 7 | 396 | 0.108% |
| | Synchronization system calls | 9 | 15387 | 84.820% |
| | Signal system calls | 3 | 79 | 0.009% |
| | Socket system calls | 8 | 12707 | 2.060% |
| | File object system calls | 2 | 56 | 0.009% |
| OS related | | 38 | 28704 | 87.040% |
| Metadata related | | 18 | 28071 | 5.270% |
| User data related | | 8 | 20339 | 7.180% |

Table 5: Total number, aggregated frequency of system calls and percentage of total execution time for the diff viewer use-case.

Table 5 shows the detected number and aggregated frequencies of different system call classes. The experimental results clearly show that the diff viewer application is a workload of mixed nature. The OS-related and metadata-related system calls are the dominating classes of system calls. The whole application budget (total number of calls) is distributed between: (1) OS related system calls – 36.95%, (2) metadata related system calls – 38.33%, (3) user data related system calls – 26.37%. Finally, we can see that the most frequent system calls are: (1) read() – read from a file descriptor – 19.55%, (2) recvmsg() – receive a message from a socket – 16.24%, (3) poll() – wait for some event on a file descriptor – 12.99%, (4)

stat() – get file status – 12.32%, (5) open() – open and possibly create a file – 9.48% (see Figure 3).

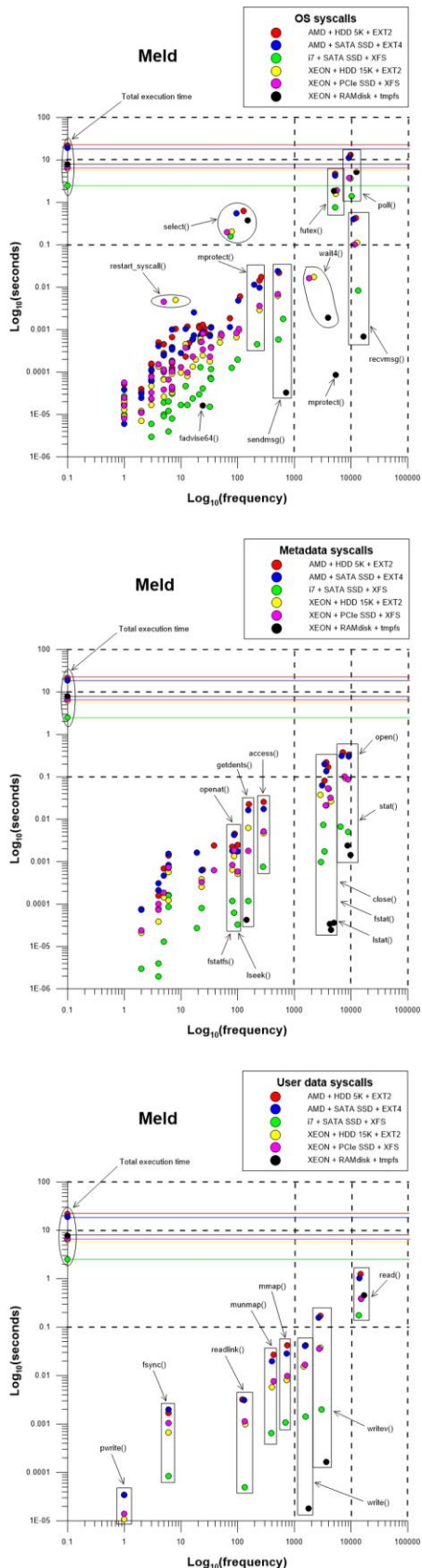

Figure 3: Diff viewer use-case.

Table 5 shows the distribution of total execution time between different system call classes. One possible conclusion is that threads synchronization dominates the application activity (84.82%). However, observing Figure 3 it becomes clearly visible that the most greedy consumers of execution time are poll() and futex() system calls. Excluding sleep time changes the whole picture and the whole application's execution time budget appears to be distributed between: (1) OS related system calls – 52.08%, (2) metadata related system calls – 19.27%, (3) user data related system calls – 28.64%. Actually, it is possible to see that diff viewer use-case is purely OS intensive application. And the most time-consuming system calls are: (1) read() – read from a file descriptor – 24%, (2) poll() – wait for some event on a file descriptor – 21.8%, (3) select() - waiting until one or more of the file descriptors become "ready" for some class of I/O operation – 12.5%, (4) futex() - fast user-space locking – 8.97%. The comparison of results for different platforms (see Figure 3) shows very interesting and unexpected results. Metadata-related system calls are executed faster for the case of a RAMdisk-based platform. However, the read() system call doesn't look any better in that case. The poll(), futex(), wait4() and read() system calls appear with the highest frequencies. Also these system calls turn out more inefficient for the RAMdisk platform. Generally speaking, these system calls are responsible for the inefficiency on the RAMdisk platform. Unexpectedly, synchronization related system calls could play an even more crucial role for the case of memory-centric applications. These system calls degrade the performance of memory-centric applications significantly in the POSIX-based environment.

### C. Internet Browser

The internet browser is very interesting and widely deployed use-case. The strace output revealed that Firefox is a multi-threaded application with 85 threads. Table 6 shows the detected number and aggregated frequencies of different system call classes. The experimental results clearly show that the internet browser application is OS intensive. The OS-related system calls are the dominating class of system calls. The whole application budget (total number of calls) is distributed between: (1) OS related system calls – 74.5%, (2) metadata related system calls – 10.83%, (3) user data related system calls – 14.19%.

| | System call Class | System calls Total | Aggregated Frequency | Execution Time (%) |
|---|---|---|---|---|
| | Process/Thread system calls | 22 | 657 | 0.0050% |
| | Memory system calls | 7 | 12606 | 0.0640% |
| | Synchronization system calls | 11 | 107519 | 93.0950% |
| | Signal system calls | 4 | 53 | 0.0002% |
| | Socket system calls | 18 | 93374 | 6.1024% |
| | File object system calls | 4 | 7539 | 0.0367% |
| OS related | | 66 | 221748 | 99.3033% |
| Metadata related | | 21 | 32741 | 0.2599% |
| User data related | | 10 | 42747 | 0.4356% |

Table 6: Total number, aggregated frequency of system calls and percentage of total execution time for the internet browser use-case.

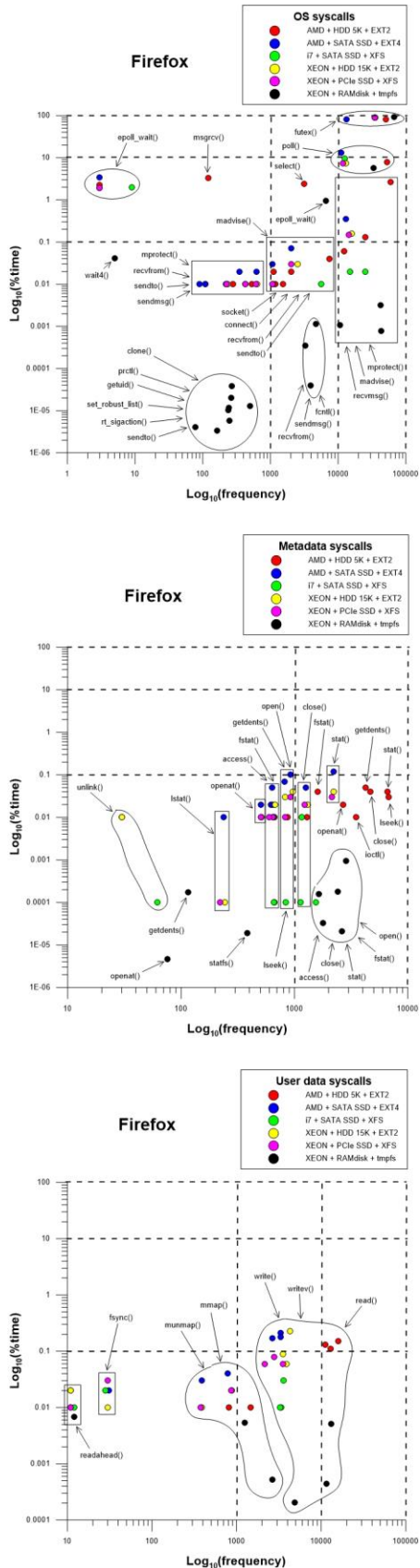

Figure 4: Internet browser use-case.

Finally, the most frequent system calls are: (1) recvmsg() - receive a message from a socket – 23.21%, (2) poll() - wait for some event on a file descriptor – 20.72%, (3) futex() - fast user-space locking – 19.99%, (4) recvfrom() - receive a message from a socket – 9.87%, (5) write() - write to a file descriptor – 6.18%, (6) writev() - write data into multiple buffers – 5.06%, (7) read() - read from a file descriptor – 4.34% (see Figure 4). Table 6 shows the distribution of total execution time between different system call classes. One conclusion is that the dominating component of the application activity comes from threads synchronization (93.09%). However, Figure 4 makes plain that the most greedy consumers of execution time are the poll() and futex() system calls. The whole application's execution time budget after the exclusion of sleep time is distributed between: (1) OS related system calls – 80.48%, (2) metadata related system calls – 6.35%, (3) user data related system calls – 13.15%. And the most time-consuming system calls are: (1) futex() - fast user-space locking – 50.3%, (2) poll() - wait for some event on a file descriptor – 24.9%, (3) recvmsg() - receive a message from a socket – 8.18%. The comparison of results for different platforms (see Figure 4) shows very interesting results. The futex(), poll(), epoll_wait(), read() are the frequent system calls in the Firefox use-case. These system calls define the main overhead of the application. One can see that the RAMdisk platform is unable to improve the performance of the Firefox use-case. Unexpectedly, the readahead() system call is even more inefficient for the case of the RAMdisk platform. The RAMdisk platform does improve the performance of metadata and user data related system calls. However, OS related system calls are much more time-consuming and synchronization related system calls play the most unpredictable role in the whole application runtime.

### D.  OpenOffice

The OpenOffice is the typical case of a widely used package of applications including a word processor, spreadsheet engine and other applications. The strace output revealed that OpenOffice Calc is multi-threaded application with 57 threads.

| System call Class | System calls Total | Aggregated Frequency | Execution Time (%) |
|---|---|---|---|
| Process/Thread system calls | 16 | 400 | 0.055% |
| Memory system calls | 7 | 960 | 0.448% |
| Synchronization system calls | 9 | 9317 | 85.064% |
| Signal system calls | 4 | 68 | 0.006% |
| Socket system calls | 15 | 8975 | 9.359% |
| File object system calls | 4 | 224 | 0.023% |
| OS related | 55 | 19944 | 94.957% |
| Metadata related | 22 | 20698 | 3.408% |
| User data related | 10 | 7366 | 1.633% |

Table 7: Total number and aggregated frequency of system calls and percentage of total execution time for the OpenOffice Calc use-case.

Table 7 shows the detected number and aggregated frequencies of different system call classes. The experimental results show that the OpenOffice Calc application is OS + metadata intensive. The OS-related and metadata-related system calls are the dominating classes of system calls. The whole application budget (total number of calls) is distributed between: (1) OS related system calls – 40.99%, (2) metadata related system calls – 43%, (3) user data related system calls – 15.32%.

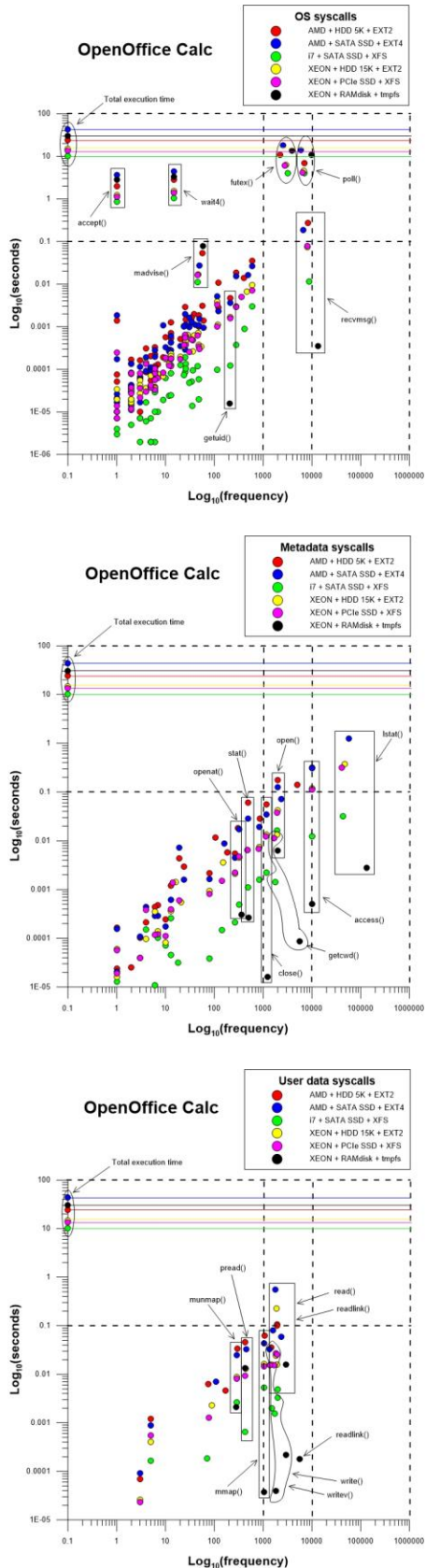

Figure 5: OpenOffice Calc use-case.

Finally, we show that the most frequent system calls are: (1) access() - check user's permissions for a file − 22.19%, (2) recvmsg() - receive a message from a socket − 18.33%, (3) poll() - wait for some event on a file descriptor − 15.32%, (4) lstat() - get file status − 11.1%, (5) futex() - fast user-space locking − 4.84%, (6) open() - open and possibly create a file − 4.36%, (7) read() - read from a file descriptor − 4.24% (see Figure 5). Table 7 shows the distribution of total execution time between different system call classes. One conclusion is that threads synchronization dominates the application's activity (85.064%). However, Figure 5 shows that the most demanding consumers of execution time are poll(), futex() and wait4() system calls. The whole application's execution time budget after the exclusion of sleep time is distributed between: (1) OS related system calls − 40.69%, (2) metadata related system calls − 48.41%, (3) user data related system calls − 10.88%. And the most time-consuming system calls are: (1) lstat() - get file status − 32.2%, (2) poll() - wait for some event on a file descriptor − 24.1%, (3) access() - check user's permissions for a file − 11.3%, (4) recvmsg() - receive a message from a socket − 7.83%, (5) futex() - fast user-space locking − 7.02%. The comparison of results for different platforms (see Figure 5) shows very interesting results. One can see that the efficiency of metadata related system calls can be improved dramatically for the case of the RAMdisk based platform. But read() and pread() system calls are the main performance bottlenecks of user data related system calls. The key bottlenecks of the application are poll(), futex(), wait4() and accept() system calls. These system calls increase the whole application overhead dramatically for the case of the RAMdisk based platform.

### E. Compilation

The compilation is very interesting workload with intensive access for both metadata and user data. The strace output revealed that the make utility is able to compile with up to 475 threads. Table 8 shows the detected number and aggregated frequencies of different system call classes. The experimental results clearly show that the make utility is a metadata intensive use-case. The whole application budget (total number of calls) is distributed between: (1) OS related system calls − 22.8%, (2) metadata related system calls − 63.09%, (3) user data related system calls − 13.21%.

| | System call Class | System calls Total | Aggregated Frequency | Execution Time (%) |
|---|---|---|---|---|
| | Process/Thread system calls | 17 | 3771 | 0.143% |
| | Memory system calls | 2 | 14950 | 0.851% |
| | Synchronization system calls | 3 | 770 | 92.820% |
| | Signal system calls | 3 | 9401 | 0.304% |
| | Socket system calls | 2 | 257 | 0.020% |
| | File object system calls | 5 | 1040 | 0.043% |
| OS related | | 32 | 30189 | 94.184% |
| Metadata related | | 23 | 81277 | 4.407% |
| User data related | | 5 | 17031 | 1.407% |

Table 8: Total number, aggregated frequency of system calls and percentage of total execution time for the make utility use-case.

Finally, we can see that the most frequent system calls are: (1) lstat() - get file status − 33.13%, (2) open() - open and possibly create a file − 13.48%, (3) brk() - change data segment size − 10.05%, (4) read() - read from a file descriptor − 5.83% (see Figure 6). Table 8 shows the distribution of total execution time between different system call classes. One conclusion is that the dominating component of the application activity is threads synchronization (92.82%).

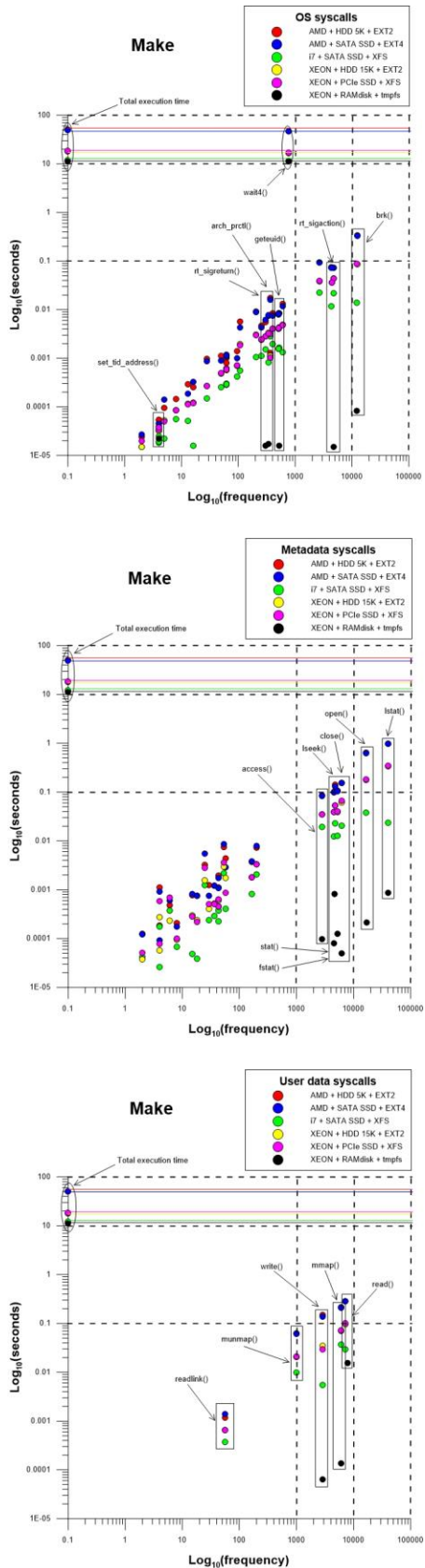

Figure 6: The make utility use-case.

However, if anyone takes a look on Figure 6 makes obvious that the highest consumer of execution time is the wait4()

system call. The whole application's execution time budget after the exclusion of sleep time is distributed between: (1) OS related system calls – 21.33%, (2) metadata related system calls – 61.16%, (3) user data related system calls – 17.49%. And the most time-consuming system calls are: (1) lstat() - get file status – 27.2%, (2) open() - open and possibly create a file – 14.4%, (3) read() - read from a file descriptor – 7.88%, (4) brk() - change data segment size – 6.87%. The RAMdisk based platform works really efficiently in the case of the make utility. All types of system calls were improved by the use of the RAMdisk based platform. However, the wait4() system call has practically unchangeable overhead that dominates the total execution time. One of the critical difference is the CLI based interface of the application. It is very probable that GUI libraries add significant overhead and contribute to make an application more inefficient.

## VI. System Calls in POSIX-based OS

### A. The Most Frequent System Calls

Table 9 shows the most frequent system calls in the use-cases we investigated.

| Use-case | Frequency | | |
|---|---|---|---|
| | OS syscalls | Metadata syscalls | User data syscalls |
| Software installation | poll() – 5.72% | stat() – 87.33% | |
| Meld (diff viewer) | recvmsg() – 16.24%<br>poll() – 12.99% | stat() – 12.32% | read() – 19.55% |
| Firefox | recvmsg() – 23.21%<br>poll() – 20.72%<br>futex() – 19.99%<br>recvfrom() – 9.87% | | |
| OpenOffice Calc | recvmsg() – 18.33%<br>poll() – 15.32% | access() – 22.19%<br>lstat() – 11.1% | |
| Autoreconf | | stat() – 17.73%<br>close() – 11.81% | read() – 16.37%<br>mmap() – 10.05% |
| Configure | rt_sigprocmask() – 15.39%<br>rt_sigaction() – 12.48% | lstat() – 13.74% | |
| Make | brk() – 10.05% | lstat() – 33.13%<br>open() – 13.48% | |

Table 9: The most frequent system calls.

One can see that metadata and OS related system calls are the most important classes for all use-cases. All applications frequently get file status (stat(), lstat()), check user's permissions (access()), open/close files (open(), close()). The most important OS-related system calls are: (1) synchronization related system calls (poll(), futex()), (2) socket related system calls (recvmsg(), recvfrom()), (3) signal related system calls (rt_sigprocmask(), rt_sigaction()).

### B. The Most Time-Consuming System Calls

Table 10 shows the most time-consuming system calls in the use-cases we investigated. The synchronization related system calls (poll(), futex(), wait4()) are the most time-consuming ones. Some applications use such system calls very frequently (for example, Meld, Firefox, OpenOffice). But even when the frequency of synchronization related system calls is not dominating this type of system calls are the main cause of time-consumption in the application (see Figure 1 - 6). This can be seen in the peculiar histogram of synchronization related system calls (for example, see Figure 2). Every histogram has really wide timings distribution that, finally, determines the total execution time of an application. Also the really important and unexpected conclusion is that even RAMdisk-based platform is unable to solve the issue with synchronization related system calls.

| Use-case | Time-consumption | | |
|---|---|---|---|
| | OS syscalls | Metadata syscalls | User data syscalls |
| Software installation | poll(), futex(), restart_syscall() | stat() | read() |
| Meld (diff viewer) | poll(), futex(), select(), recvmsg() | open(), close(), stat(), fstat(), lstat() | read() |
| Firefox | futex(), poll(), epoll_wait(), recvmsg() | open(), close(), stat(), fstat() | read() |
| OpenOffice Calc | futex(), poll(), wait4(), accept(), recvmsg() | lstat(), stat(), access(), open(), close() | read(), readlink() |
| Autoreconf | wait4(), mprotect(), brk(), rt_sigaction() | stat(), fstat(), open(), close() | read() |
| Configure | wait4(), rt_sigprocmask(), rt_sigaction(),mprotect(), brk() | stat(), fstat(), lstat(), open(), close(), access() | read() |
| Make | wait4(), rt_sigprocmask(), rt_sigaction(),mprotect(), brk() | stat(), fstat(), lstat(), open(), close(), access() | read() |

Table 10: The most time-consuming system calls.

We detected some outliers (restart_syscall(), select(), epoll_wait(), accept()) that have really low frequency and very unreasonable time consumption. Finally, it became clear that the issue with such system calls cannot be resolved even for the case of the RAMdisk-based platform. The rest of the Table 10 confirms that socket related system calls (recvmsg(), recvfrom()) and signal related system calls (rt_sigprocmask(), rt_sigaction()) are responsible for significant part of the application's time consumption. Metadata system calls compete in time-consumption with user data system calls (read(), for example). Sometimes metadata-related system calls (for example, stat()) can be significantly much time-consuming than all user data system calls (for example, in the software installation use-case). One can conclude that OS and metadata related system calls are the most important area of optimization for the case of NVM/SCM memory environment.

### C. RAMdisk Case

The RAMdisk platform is one of the vehicle used to simulate the behavior of POSIX-based OSes in NVM/SCM memory environment. Figure 1 - 6 show that the RAMdisk platform decreases the aggregated timing for practically all types of system calls. However, this platform has revealed critical bottlenecks of POSIX and CPU-centric paradigms. Unexpectedly, the application's total execution time can be even greater for the case of the RAMdisk based platform. The key reason is that the synchronization overhead remains practically unchanged. For example, the aggregated timing of poll(), futex(), wait4() system calls is greater in the case of the OpenOffice Calc application on the RAMdisk platform (see Figure 5). The CPU-centric computing paradigm can be the basis for such issue. Because, context switches, task scheduling and competition for main memory between CPU cores becomes more obvious and critical in the case of the RAMdisk platform. It is possible to distinguish the tendency that the total execution time is defined mostly by CPU architecture. Changing the type of persistent storage device may improve the total execution time slightly but the CPU architecture remains responsible for real improvement to the computer system's performance. Finally, we conclude that CPU-centric and POSIX-based paradigms are the key bottlenecks in NVM/SCM memory systems.

## VII. POSIX OS IN NVM/SCM ENVIRONMENT

### A. Critical Properties of POSIX-based OSes

Synchronization primitives play a very critical role in POSIX-based OSes. The frequency of this type of system calls is very high for all use-cases. Also a dominating portion (50% - 90%) of applications' execution time is spent in synchronization primitives. One can conclude that any multi-threaded application spends significant amount of time waiting to access the shared resources. Also, the POSIX-based paradigm relies on synchronization primitives in a multi-threaded environment. Generally speaking, this implies that fundamental abstractions of POSIX-based OS are the reason for such critical bottlenecks. Abstractions of file and thread play cornerstone role in POSIX-based OSes. A file is usually defined as an "infinite" persistent byte stream. A process/thread can be imagined as the fundamental abstraction of elementary OS task. The investigated use-cases showed that the major portion of the whole application's execution time is spent in poll(), futex() and wait4() system calls. Even the RAMdisk based platform cannot resolve the bottleneck caused by synchronization primitives. The poll() system call is designed to wait for some event on a file descriptor. It is possible to expect that if all data is kept in DRAM then the overhead of the poll() system call should be lower. Because, usually, if requested data is not in the page cache then it needs to be read from the persistent volume into DRAM. As a result, the RAMdisk platform can exclude such copy operations because everything remains in DRAM. However, the measurement results don't show any decrease in overhead in the case of poll() system call on the RAMdisk platform. The most probable reason is that the page cache approach is reasonably efficient and, as a result, the aggregated time budget of the poll() system call is practically the same for all investigated hardware platforms. Generally speaking, the OS keeps file's content in the radix tree that contains the memory pages retrieved from the persistent volume. If some application tries to access the portion of file that is not in the page cache then the requested set of physical sectors have to be retrieved from the persistent volume. However, the poll() system call is used for the whole file. This implies that if a requested file's content is in the page cache then the interaction with persistent storage cannot affect the poll() system call and there is no visible difference between RAMdisk and any other platform. Finally, the task scheduler plays a key role as the subsystem that distributes the CPU's time slices between threads/processes. It means that the coarse granularity (the poll() system call locks the whole file) is the reason why the task scheduler plays the main role in POSIX-based OSes. The task scheduler nature makes the situation with the poll() system call non-deterministic, unpredictable and unmanageable. File abstraction is able to be critical issue for the case of poll() system call in the environment of NVM/SCM memory. The file concept creates the non-transparent binary borders that hide the granularity of the file's internal data. Finally, namely file's internal data needs in synchronization of accesses from different threads but not file itself. This issue can be more critical in the case of NVM/SCM memory because a file will be always in byte-addressable, fast persistent memory and the task scheduler can introduce more overhead when dealing with "faster" threads that work with persistent memory. In the case of futex() (waiting until a certain condition becomes true) and wait4() (wait for process to change state), the role of the task scheduler is more clear. The synchronization mechanisms of futex() and wait4() are based completely on the OS's internal structures. As a result, it is hard to expect any difference between RAMdisk and other hardware platforms. The distribution of poll() system call timings (Figure 2) makes it clear that it is impossible to ignore the nature of the task scheduler. Using NVM/SCM memory in the current computing paradigm simply increases the competition for shared resources (memory and CPU time slices) and, as a result, the task scheduler overhead will be increased.

### B. Criticism of POSIX-based OSes

The task scheduler was invented because it became necessary to share CPU cores between several tasks in order to create a pre-emptive multi-tasking environment. This approach introduces the thread/process concept, virtual memory concept and context switch concept. A developer is responsible for taking the decision how to split the whole execution flow of a multi-threaded application on some set of threads. As a result, the task scheduler needs to manage various type of threads with different execution environment requirements (application's threads might have different functionality and different importance in the application context). But the task scheduler has to treat the threads in a uniform way. The threads' priority attempts to solve this problem. But for an application that contains hundreds or thousands threads distributing the priority is not a trivial problem. Moreover, if the OS needs to manage a hundreds of concurrently working applications achieving a really good and efficient task scheduling policy is a hard task. If the computing system has only one core then the illusion of "simultaneous" execution is created by pre-emptive multi-tasking approach. This means that the task scheduler distributes CPU time slices between all threads by means of a well-defined algorithm. But every task should be represented by an execution context because of load/store model of CPU. Finally, this means that the CPU has to unload the execution context of one task and to load the execution context of another task when switching context. It is possible to conclude that limited number of CPU's registers and process-centric model of POSIX-based OS is the fundamental reason of the task scheduler's inefficiency. This issue cannot be solved by simple adding NVM/SCM memory into the computing system. Another fundamental problem is the increasing number of physical CPU cores in the system. On the one hand, increasing the number of CPU cores potentially improves the computer system's performance. However, the memory coherence problem neutralizes most of these performance improvements. Every CPU core has its own L1/L2 cache and these caches contain a copy of data from the DRAM memory space that is shared between cores. Any change in L1/L2 cache needs synchronization of these changes with the main copy in the DRAM memory space. And caches coherence increases the synchronization overhead further. File and process/thread abstractions are the fundamental source of inefficiency in POSIX-based OSes. And this issue becomes more critical in the case of NVM/SCM memory environment. Another fundamental peculiarity or anachronism of POSIX-based OSes is the huge amount of metadata operations that were inherited from the model of interaction with slow persistent storage devices. The reason of such peculiarity is the model accessing of any file in a POSIX-based OS: (1) open file, (2) get file status, (3) check user permissions, (4) reposition read/write file offset, (5) read/write file content, (6) close file. Such significant amount of metadata operations is in the nature of POSIX-based OSes and this overhead cannot be eliminated even for the case of NVM/SCM memory environment. But do all metadata operations really need to be used in the case of NVM/SCM memory? Actually, existing metadata structures, on-disk layout and file system concept as a whole are relict of the era of slow persistent storage devices. First of all, there is the fundamental dichotomy and replicated representation of metadata structures on the OS side and in the file system volume (on-disk layout). Does it make sense to continue to use such fundamental dichotomy or should metadata representation be unified? It is possible to foresee that the future OS paradigm needs in single and unified metadata representation for OS internals and user data "storage". Second very important point, does it make sense to organize data in files in the case of NVM/SCM memory. The traditional file system's metadata structures (superblock, block bitmap, extent tree and so on) may turn out mostly for NVM/SCM memory. Leaving the concept of file aside would clash strongly with the POSIX-based approach. But trying to conserve the POSIX-based approach will result in fundamental inefficiencies for any future applications of NVM/SCM memory.

### VIII. CONCLUSION

This paper presents a methodology to evaluate applications' latency on top of a systems platform. We attempted to set a baseline to evaluate new systems performance when those systems may be transitioning to NVM/SCM memory. Our experiments confirmed our initial hypothesis. The existing memory stack of 'modern' OSes, like the Linux kernel is not ready to assimilate NVM/SCM memories into the memory hierarchy or at same level as main memory. It doesn't leverage the improved performance of these memories like MRAM, CBRAM, PCM or ReRAM. The memory stack still uses obsolete paradigms which lead to the underuse of the potential of NVM/SCM memory technologies. The observation of experiments demonstrates the problem that CPUs are not fully utilized (software uses an average of 84% of its CPU time in idle) due to the fact that the memory stack is not capable to fully supply the CPUs and inefficiencies of process scheduling.

Given this result, it seems logical to shift efforts towards replacing the memory stack by a cleaner more effective solution for NVM/SCM memory. This research should help to understand how to evaluate the memory stack's performance impact, how to analyze platform performance and how to make the necessary changes to reach the performance promised by NVM/SCM at the application level.

APPENDIX A

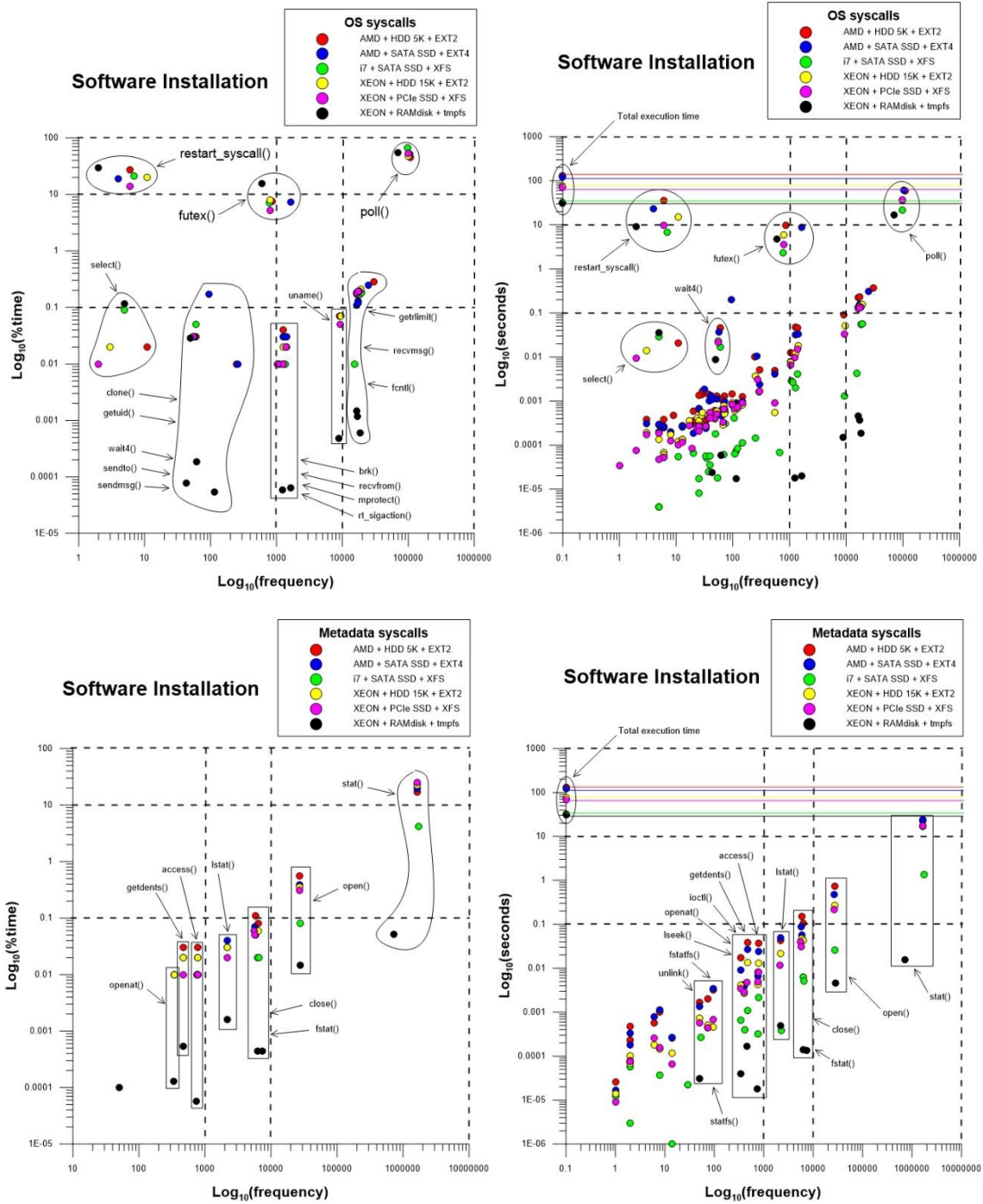

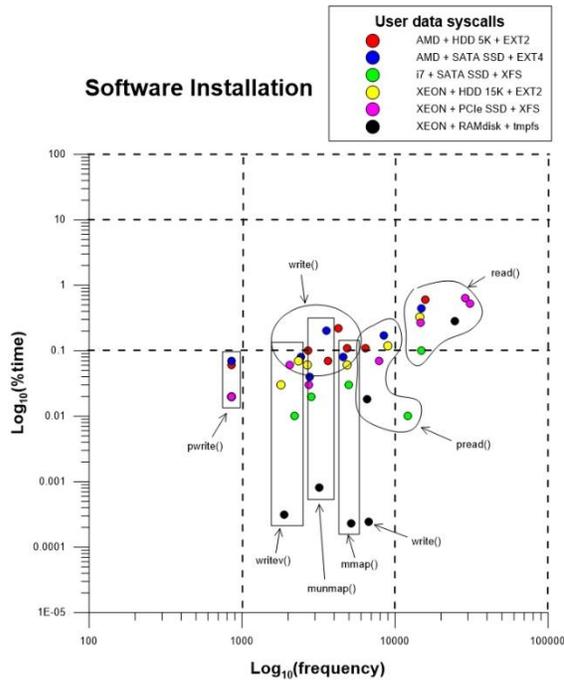

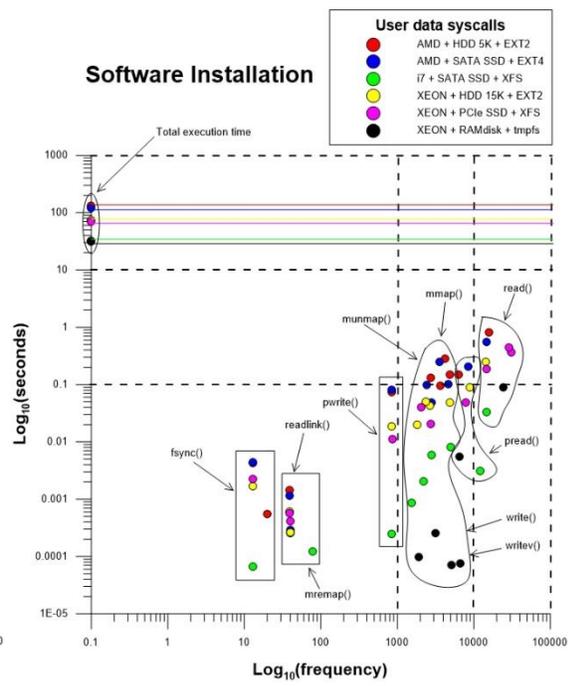

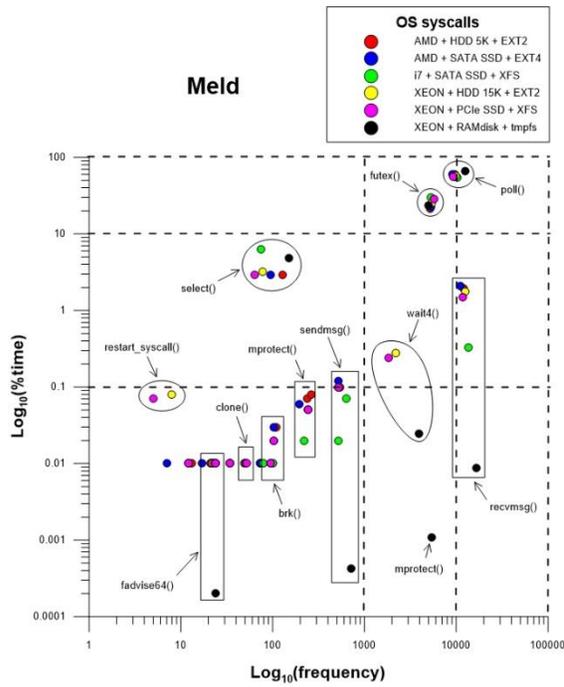

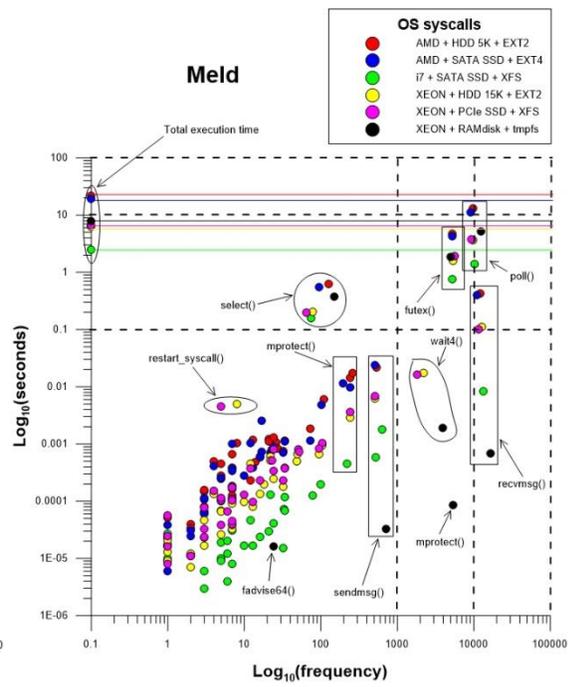

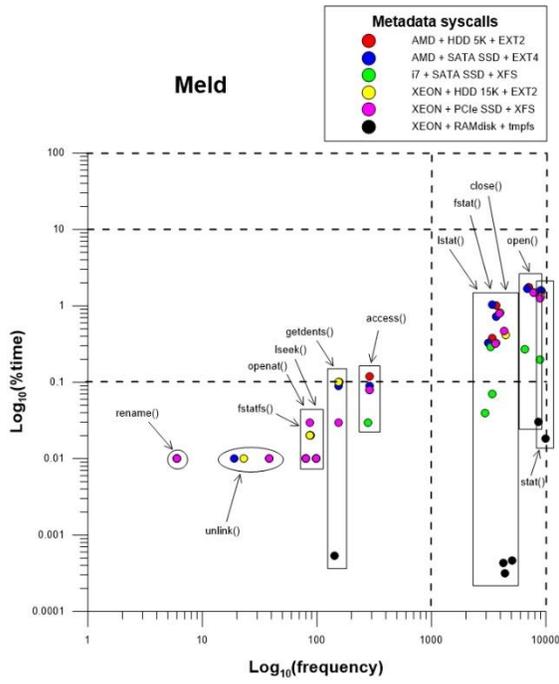

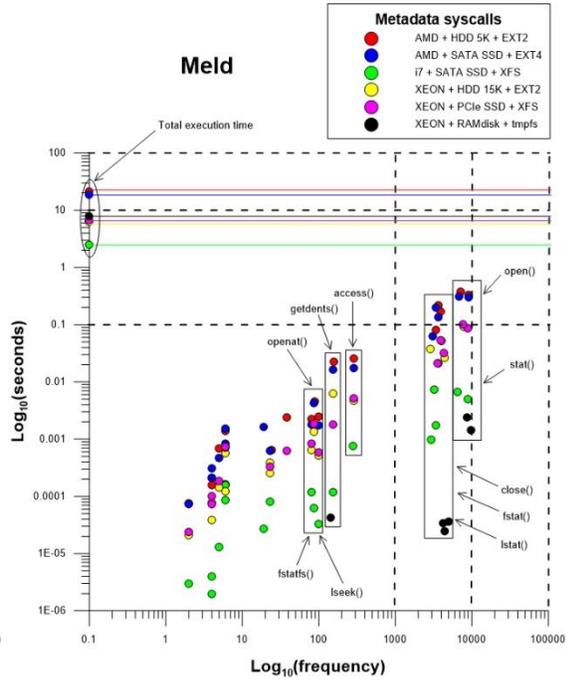

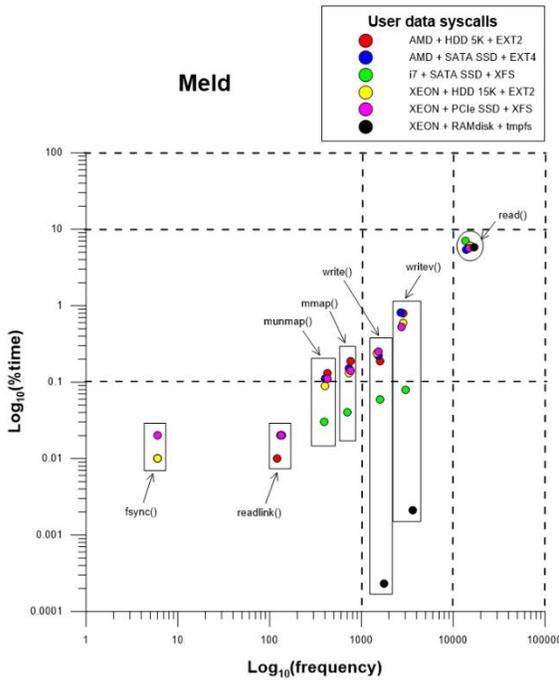

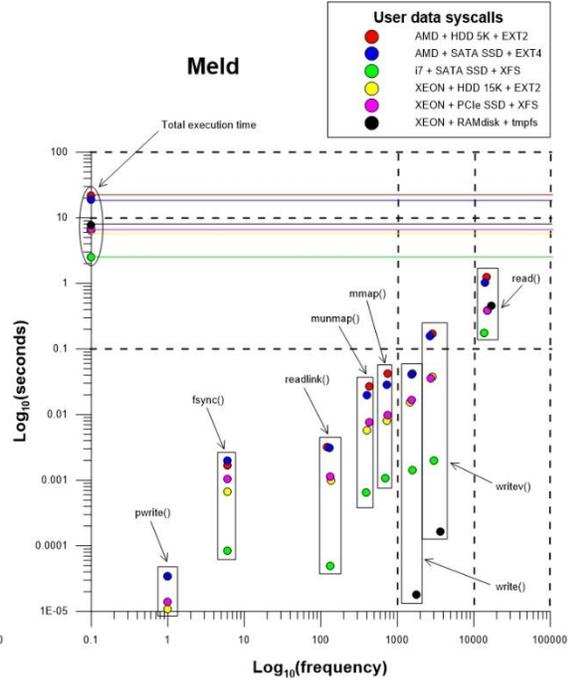

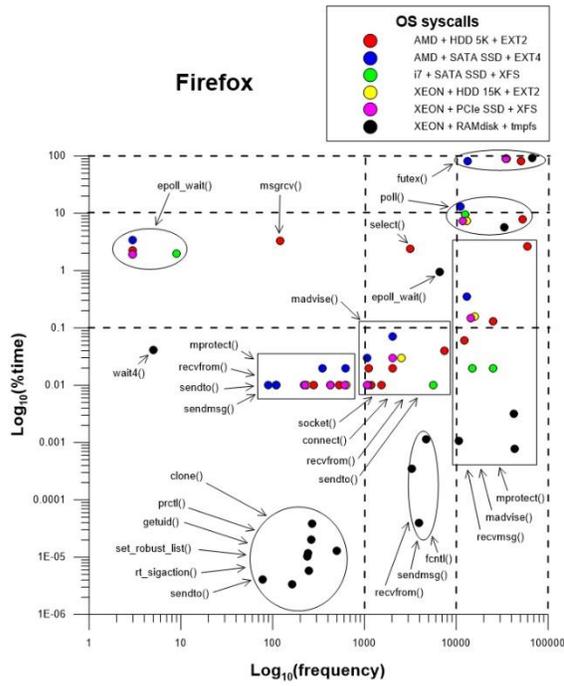

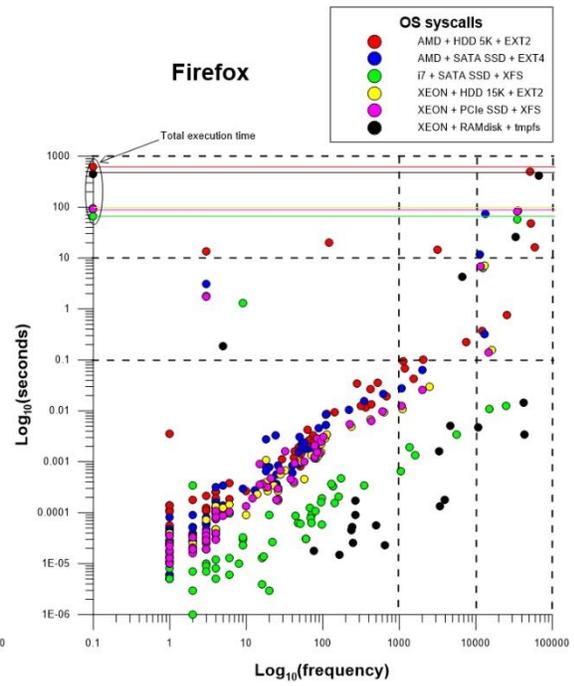

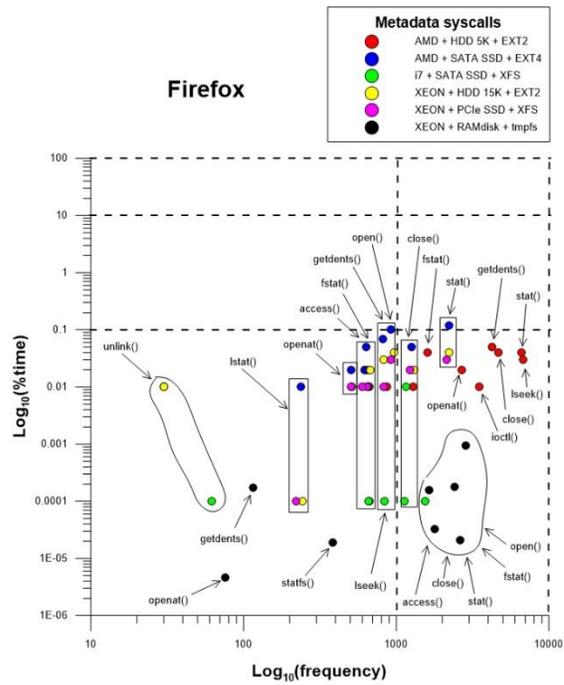

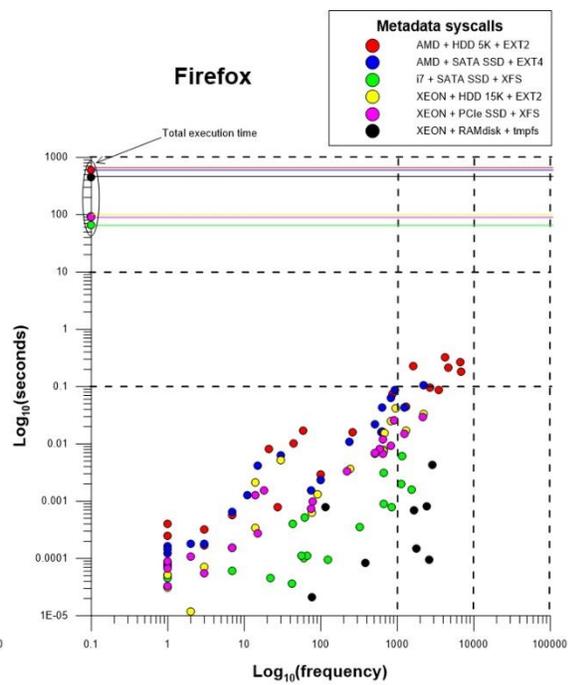

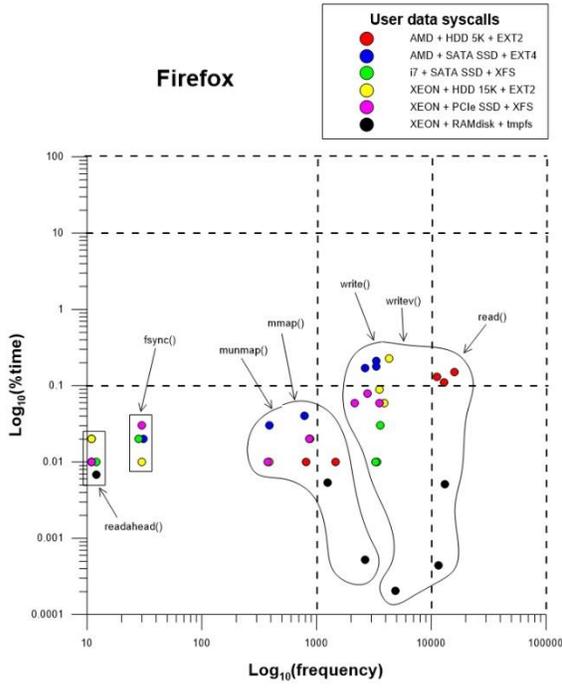

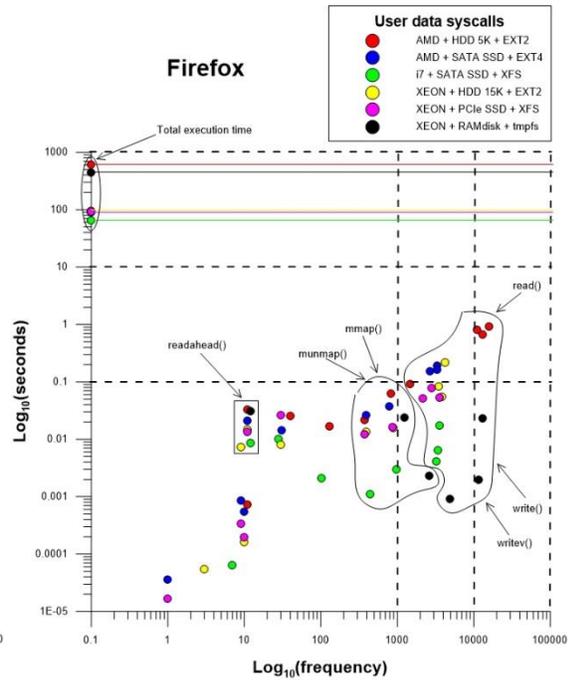

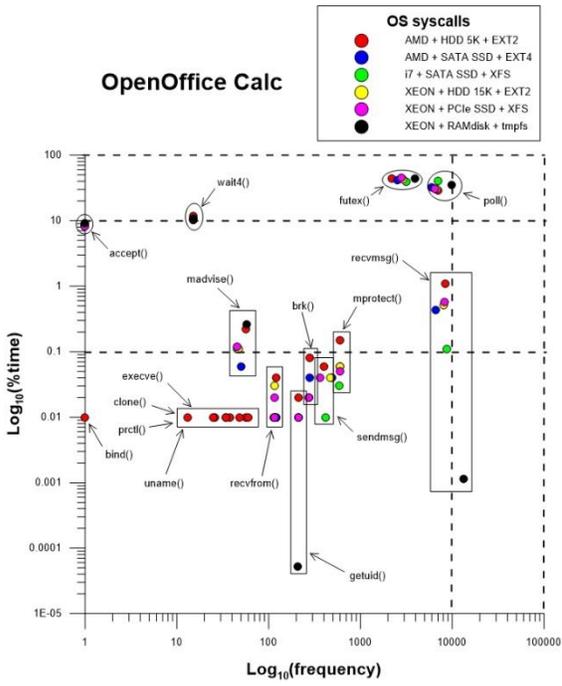

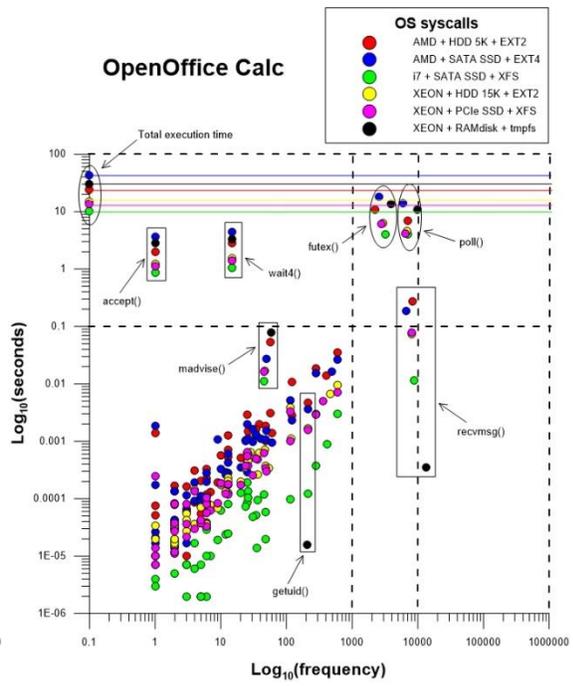

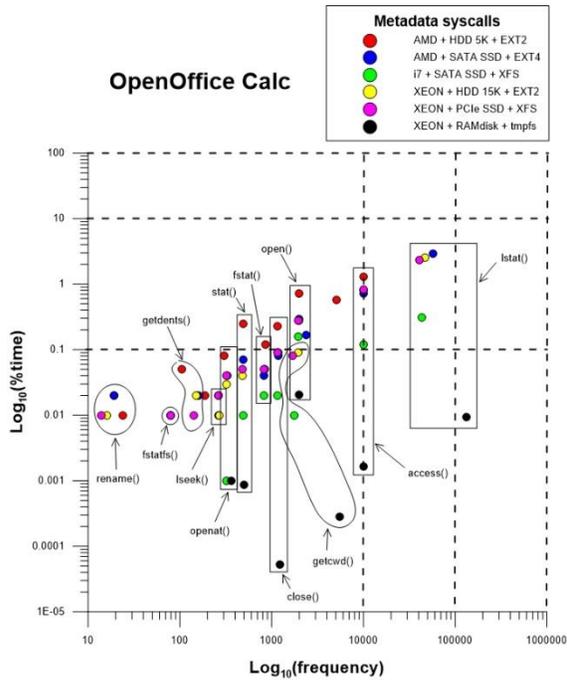

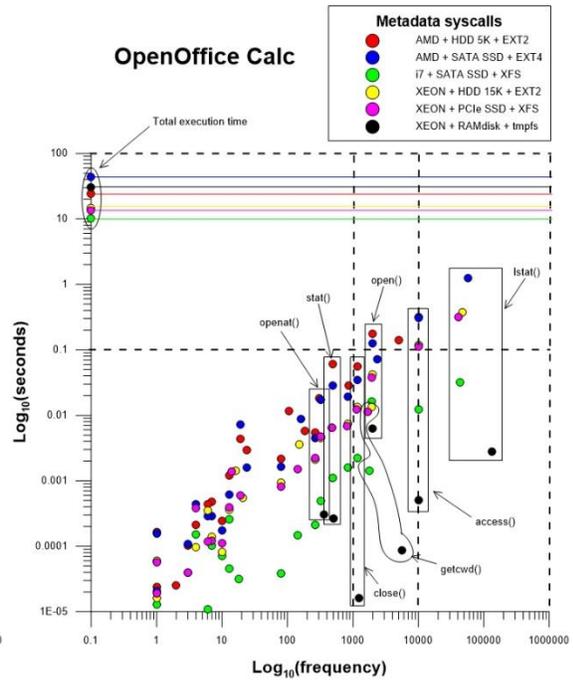

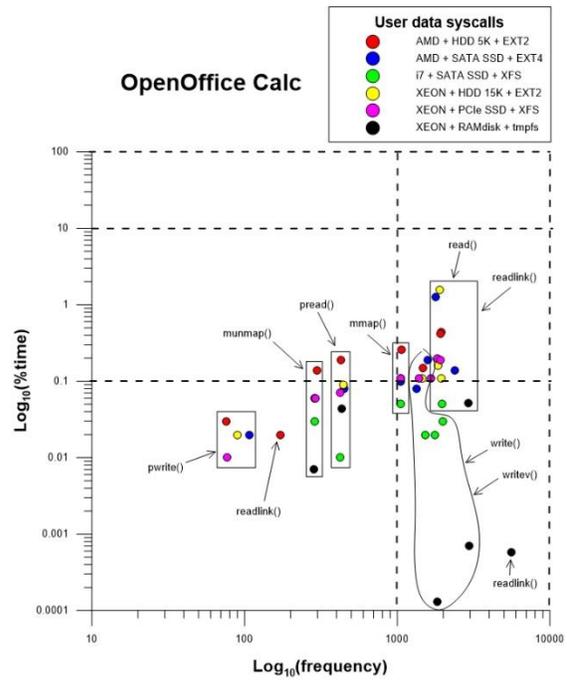

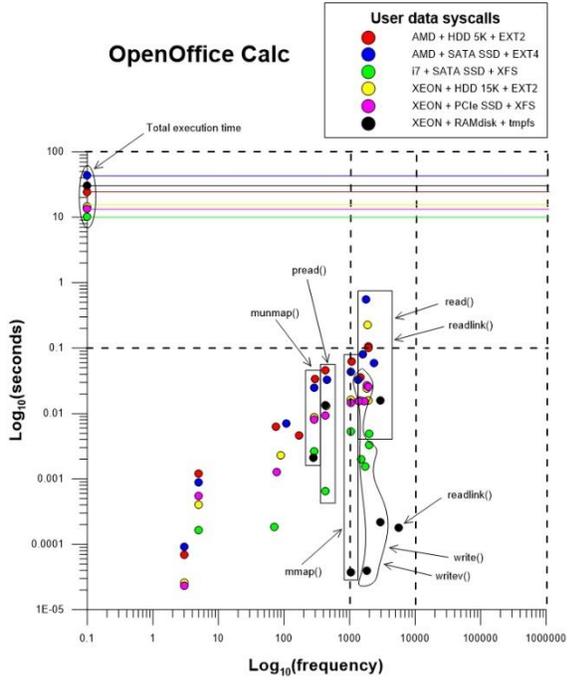

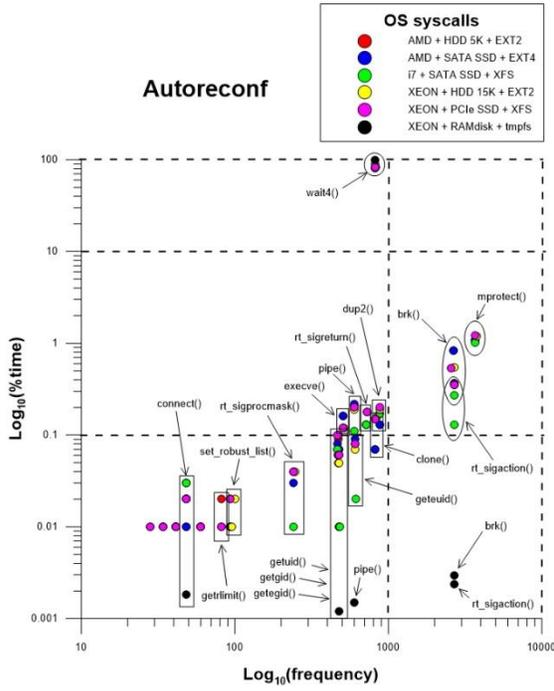

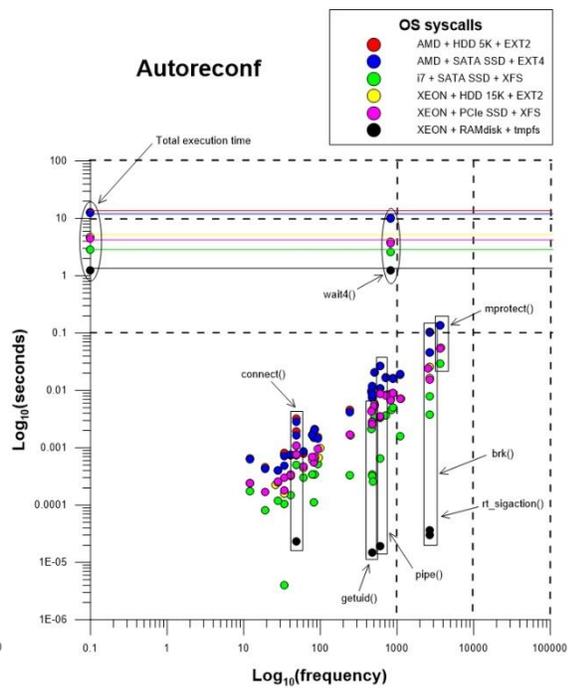

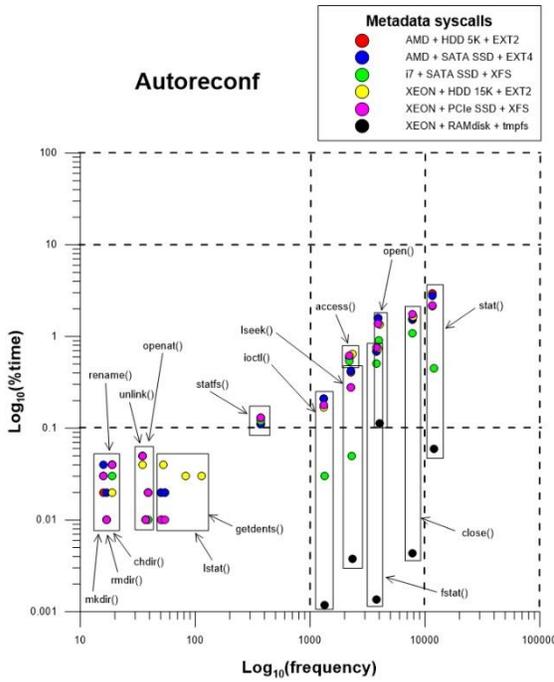

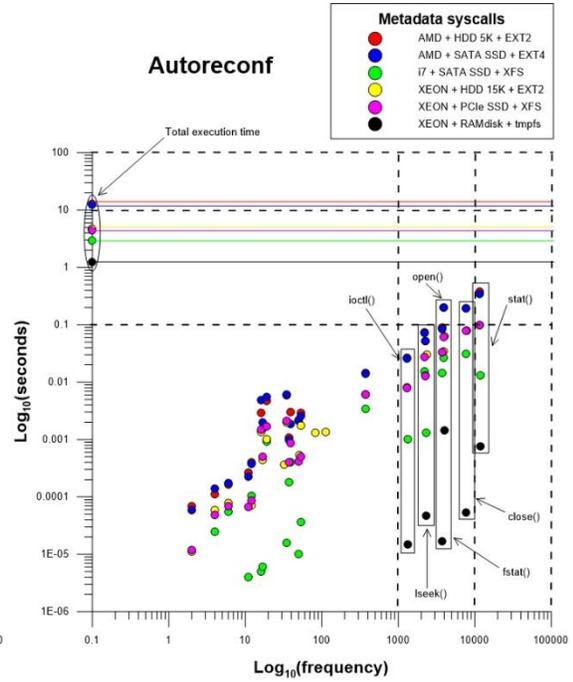

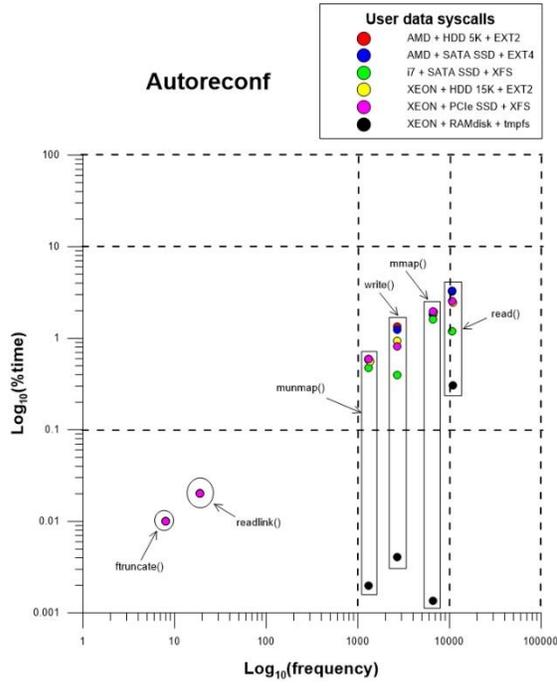

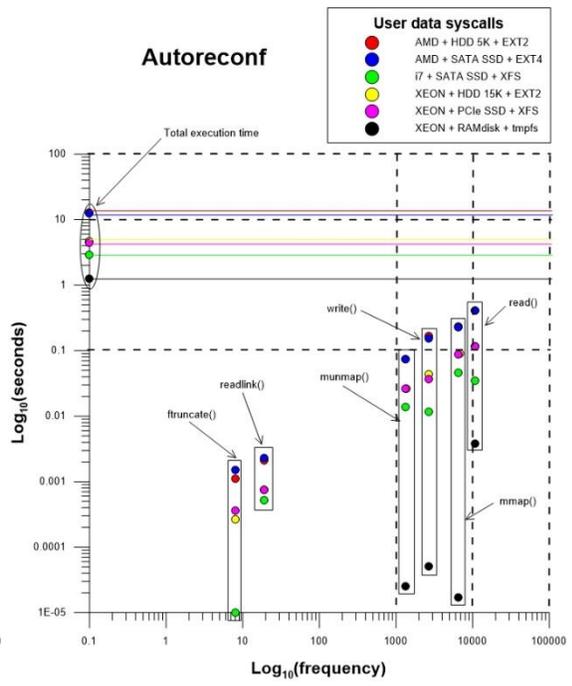

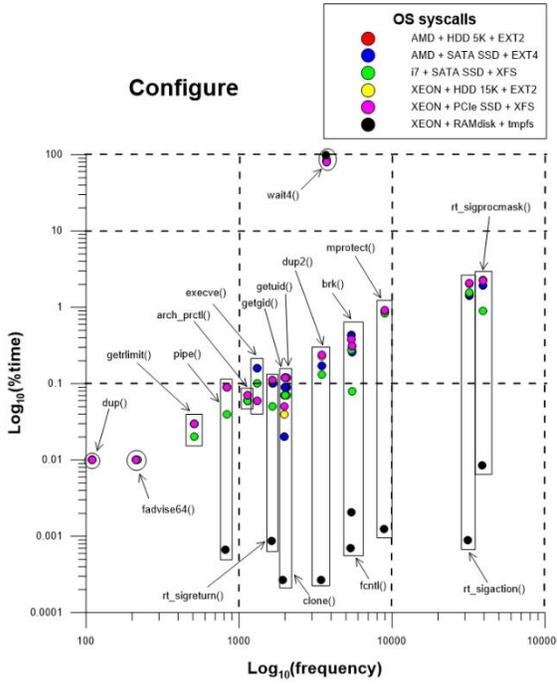

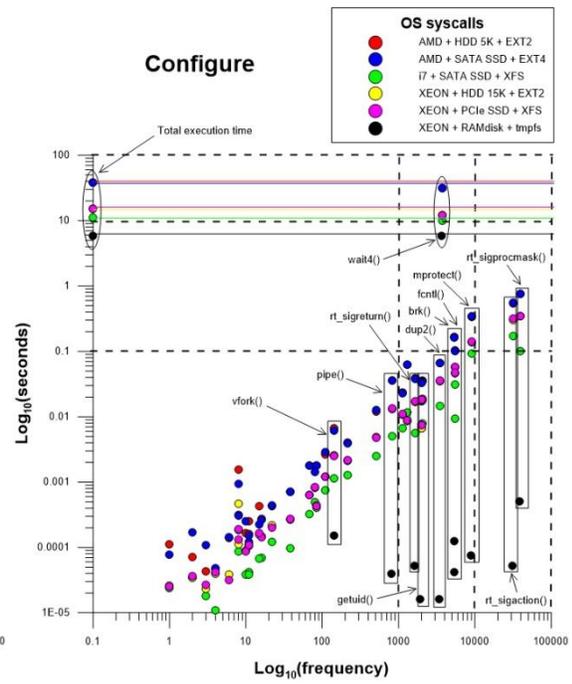

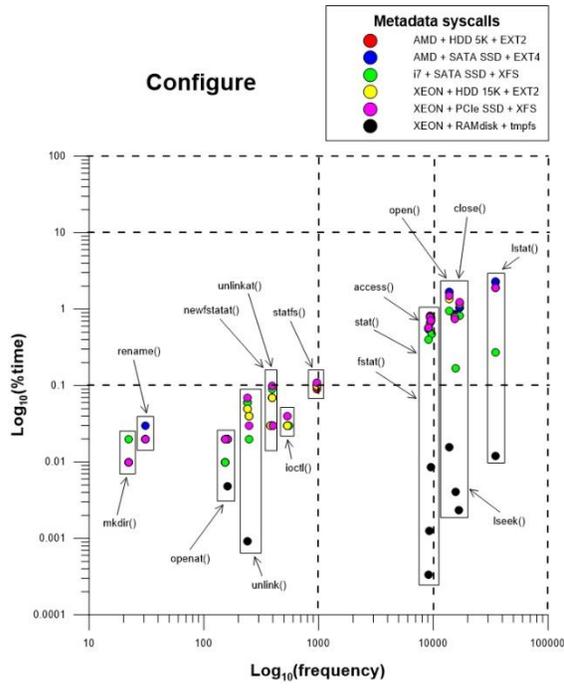

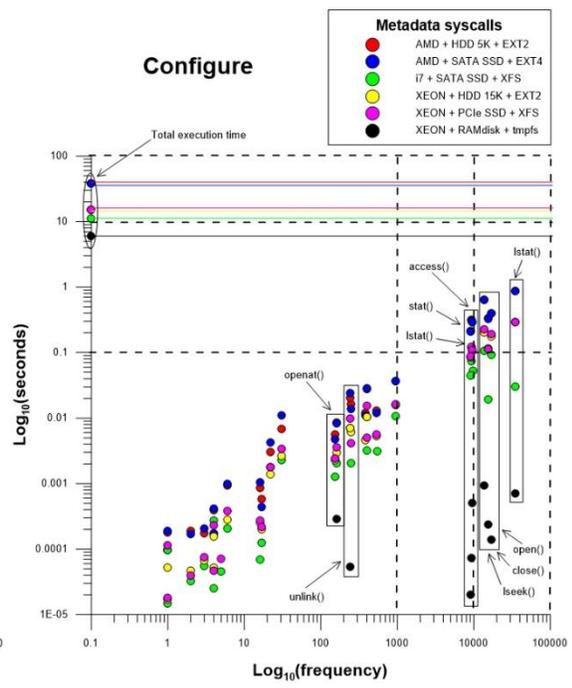

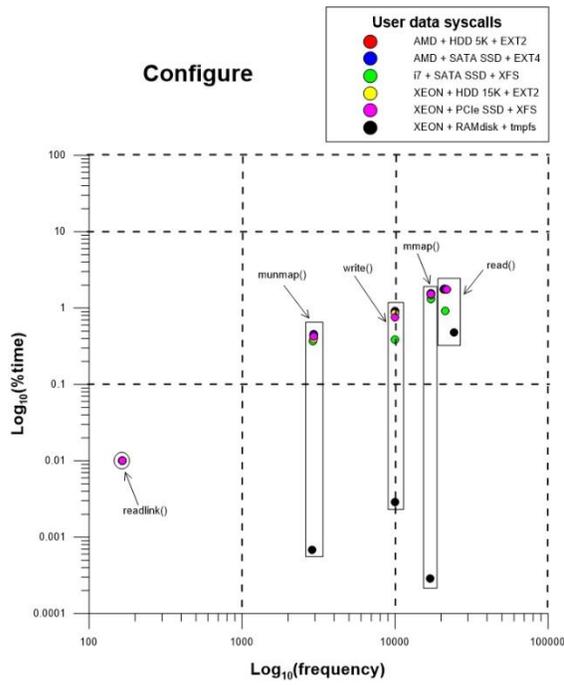

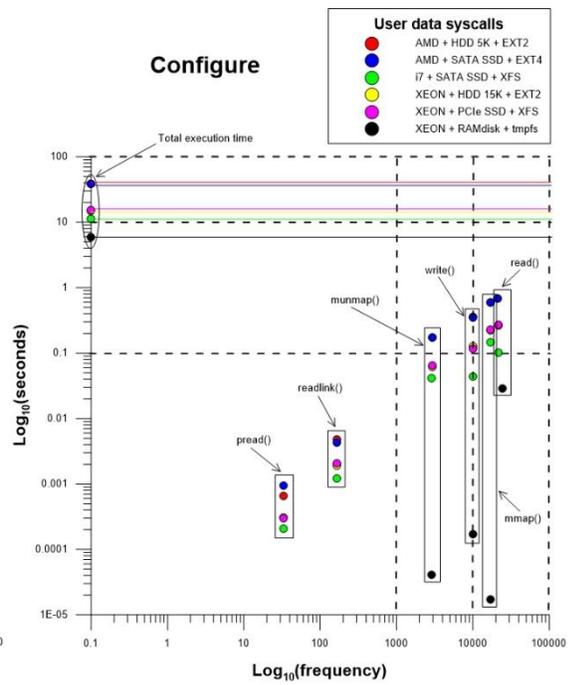

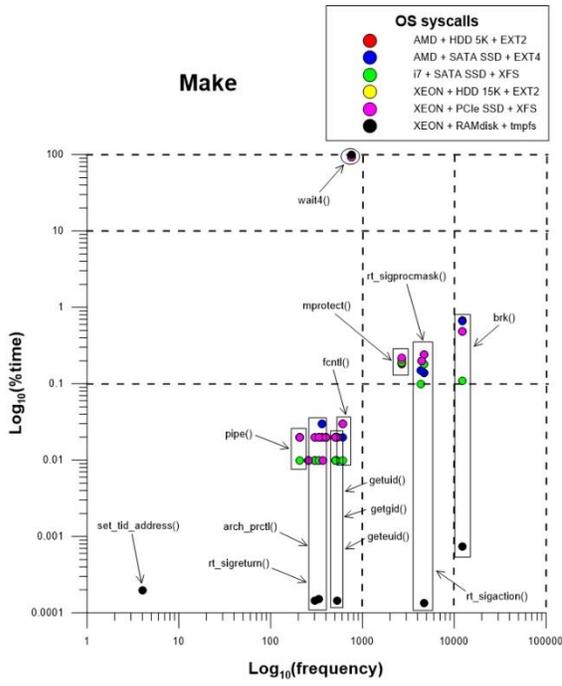

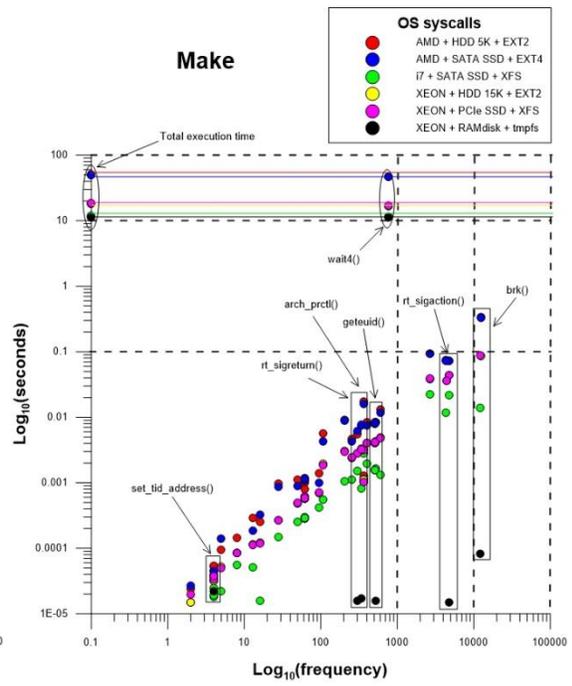

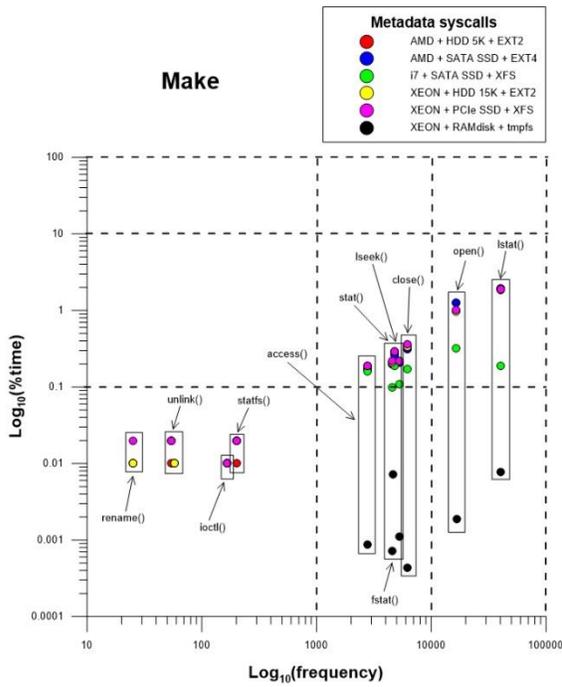

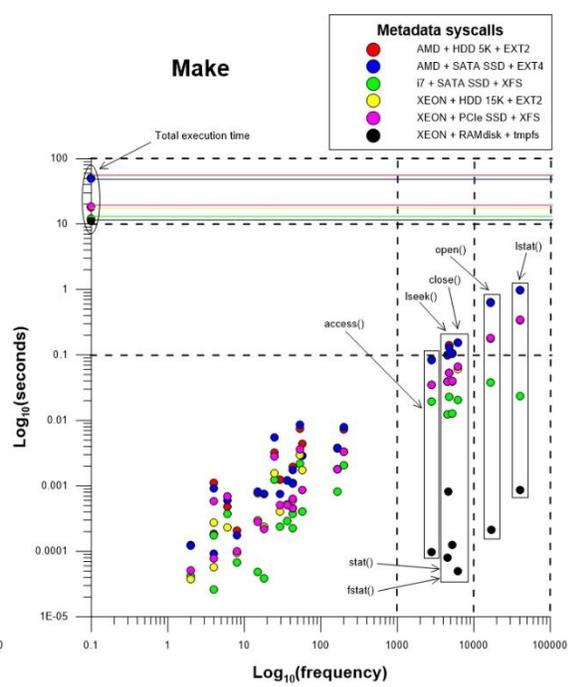

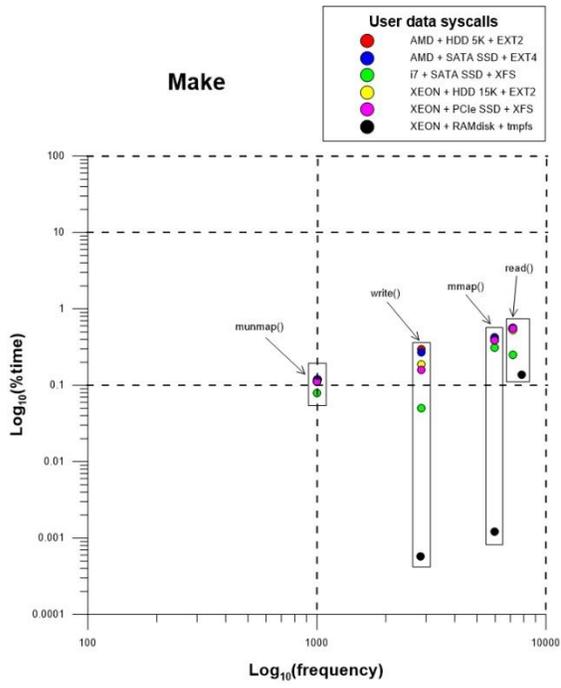

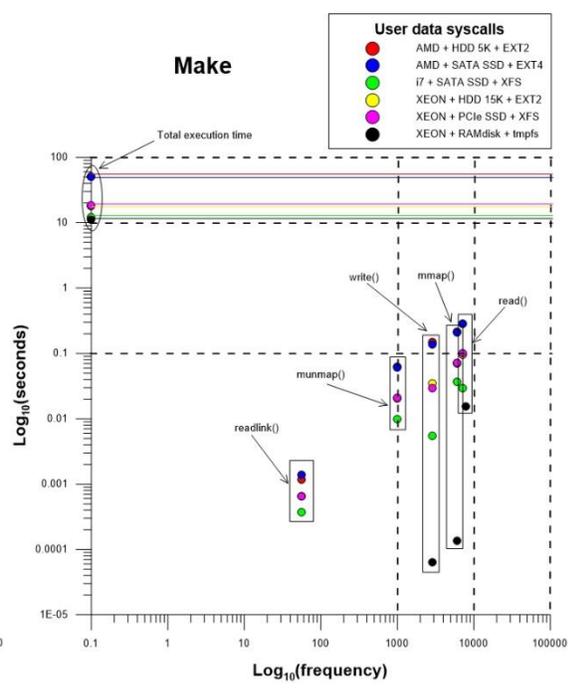